\documentclass[useamsfonts,twocolumns]{pasj00}
\draft
\usepackage{lscape}

\begin{document}
\SetRunningHead{Y. Moritani et al.}{Precessing Warped Be disk in A~$0535+262$}
\Received{xxxx/xx/xx}
\Accepted{xxxx/xx/xx}

\title{
Precessing Warped Be Disk Triggering the Giant Outbursts in 2009 and 2011 in A~$0535+262$/V725 Tau
}

\author{
Yuki \textsc{Moritani},\altaffilmark{1,2}
Daisaku \textsc{Nogami},\altaffilmark{3}
Atsuo T. \textsc{Okazaki},\altaffilmark{4}
Akira \textsc{Imada},\altaffilmark{5}
Eiji \textsc{Kambe},\altaffilmark{5}\\
Satoshi \textsc{Honda},\altaffilmark{3,6}
Osamu \textsc{Hashimoto},\altaffilmark{7}
Sahori \textsc{Mizoguchi},\altaffilmark{8,9}
Yuichi \textsc{Kanda},\altaffilmark{8,10} \\
Kozo \textsc{Sadakane},\altaffilmark{8}  and
Kohei \textsc{Ichikawa}\altaffilmark{2}
} %
\altaffiltext{1}{Hiroshima Astrophysical Science Center, Hiroshima University, 1-3-1 Kagamiyama Higashi-Hiroshima, Hiroshima 739-8526}
\altaffiltext{2}{Department of Astronomy, Faculty of Science, Kyoto University, Kitashiralawaoiwake-cho, Sakyo-ku, Kyoto 606-8502}
\altaffiltext{3}{Kwasan Observatory, Kyoto University, Yamashina-ku, Kyoto 607-8471}
\altaffiltext{4}{Faculty of Engineering, Hokkai-Gakuen University, Toyohira-ku, Sapporo 062-8605}
\altaffiltext{5}{Okayama Astrophysical Observatory, National Astronomical Observatory of Japan, Asakuchi, Okayama 719-0232}
\altaffiltext{6}{Nishi-Harima Astronomical Observatory, Center for Astronomy,University of Hyogo, 407-2, Nishigaichi, Sayo-cho, Sayo, Hyogo 679-5313}
\altaffiltext{7}{Gunma Astronomical Observatory, Takayama-mura, Gunma 377-0702}
\altaffiltext{8}{Astronomy Institute, Osaka Kyoiku University, 4-698 Asahigaoka Kashiwara, Osaka 582-8582}
\altaffiltext{9}{Sendai Astronomical Observatory, Nishikigaoka, Aoba-ku, Sendai 989-3123}
\altaffiltext{10}{Fujisoft Incorporated, Naka-ku, Yokohama, 231-8008}
\email{moritani@hiroshima-u.ac.jp}


\maketitle

\begin{abstract}
We carried out optical high-dispersion spectroscopic monitoring of the Be disk in a Be/X-ray binary A~$0535+262$/V725 Tau from 2009 to 2012, covering two giant outbursts and several normal outbursts.
This monitoring was performed in order to investigate variabilities of the Be disk due to the interaction with the neutron star in recent X-ray active phase from 2008 to 2011.
Such variabilities give a clue to uncleared detailed mechanism for very bright X-ray outbursts, which are unique to some Be/X-ray binaries with relatively wide and eccentric orbit.
In the previous letter \citep{Moritani2011}, a brief overview of line profile variabilities around the 2009 giant outburst was given and the possibility of a warped Be disk was discussed.
In this paper, a full analysis of the H$\alpha$ line profiles as well as other line profiles is carried out.
A bright blue component, or blue ``shoulder", showing up after periastron indicates the presence of a dense gas stream toward the neutron star, which is associated with observed outbursts.
We re-analyze the H$\alpha$ line profiles before 2009 (down to 2005) in order to investigate the variability of the the disk structure in the innermost region, which seems to have detached from the Be star surface by 2008.
A redshifted enhanced component is remarkable in all emission lines observed around the 2009 giant outburst, occasionally forming a triple peak.
These features indicate that the Be disk was warped in X-ray active phase.
We estimate the position of the warped region from fitting the radial velocity of the redshifted enhanced component of H$\alpha$, and find that it was very close to the periastron when two giant outbursts in 2009 and 2011 and a bright normal outburst in 2010 March occurred.
These facts strongly suggest that the warped Be disk triggered these giant outbursts.
\end{abstract}

\section{Introduction}
Be/X-ray binaries consist of a Be star and a compact object, a neutron star in general \citep{Reig2011}.
Be stars are B-type giant or dwarf stars (luminosity class III -- V) which have exhibited Balmer lines in emission at least once \citep{Porter2003}.
They have a geometrically thin circumstellar envelope called a Be disk, which forms as a result of viscous diffusion ($\lesssim$ 1 km/s) of matter ejected from an equatorial region of a rapidly rotating central star \citep{Lee1991,Carciofi2011}.

Since Be/X-ray binaries have a relatively wide ($P_{\mathrm{orb}} \; \gtrsim $ 10 days) and eccentric ($e \; \gtrsim$ 0.3) orbit in general, the mass transfer rate from the Be star to the neutron star depends on its orbital phase.
Hence, Be/X-ray binaries are generally X-ray transient sources.

The X-ray outbursts in Be/X-ray binaries are divided into two categories with respect to the luminosity: normal (type I) outbursts ($L_{\mathrm{X}} \sim 10^{36-37}$ erg s$^{-1}$) and giant (type II) outbursts ($L_{\mathrm{X}} \gtrsim 10^{37}$ erg s$^{-1}$).
The normal outbursts occur around periastron passage, and last for several days.
This type of outbursts are seen in systems with intermediate to high eccentricities where the mass transfer from the Be disk to the neutron star takes place at every periastron passage \citep{Okazaki2001b,Negueruela2001a}.
The giant outbursts, on the other hand, lasting several tens of days, are not well understood since they show no orbital modulation and occur much less frequently than normal outbursts.
Although it is widely accepted that X-ray outbursts originate from the mass transfer from the Be disk to the neutron star, detailed mechanism for X-ray outbursts is still disputable.

Recently, \citet{Okazaki2012} proposed a new scenario for both normal outbursts and giant outbursts.
They simulated the mass-accretion rate and the accretion timescale for the standard accretion flows and the radiatively inefficient accretion flows (RIAFs), where the material is transferred from the outermost part of the Be disk via a tidal stream or the Bondi-Hoyle-Lyttleton (BHL) accretion, depending on the disk size.
As a result, they found that it is the RIAF scenario which is consistent with the observational features of normal X-ray outbursts.
They also claimed that the giant outburst occurs in systems with the Be disk misaligned with the orbital plane, when the Be disk crosses the orbit of the neutron star, e.g., as a result of warping \citep{Martin2011}.

Theoretically, there are two candidate mechanisms for the warping of Be disks: the radiatively driven warping and the tidally driven warping.
\citet{Porter1998} applied the radiatively driven warping theory for accretion disks to Be disks and found that the warp is induced in an inner part of the disk by the irradiation from the central star.
He also found that the warp is more efficient for late type Be stars than for early type Be stars.
On the other hand, \citet{Martin2011} performed numerical simulations of the tidally driven warping of viscous decretion Be disks in misaligned Be/X-ray binaries.
In such systems, the tidal torque by the neutron star has a strong effect on the Be disk, making it warped and twisted except in systems with the longest orbital periods.

Observationally, the warping episode of a Be disk was suggested for several systems \citep{Hummel1998,Negueruela2001b,Reig2007}.
\citet{Hummel1998} analyzed spectacular spectral variations of $\gamma$ Cas and 59 Cyg and found that the sequence of variations from a shell-line profile to a single-peaked profile is likely caused by the change of the disk inclination.
As for the Be/X-ray binaries, long-term monitoring observations of the H$\alpha$ line profile of 4U~$0115+634$/V635 Cas by \citet{Negueruela2001b} and \citet{Reig2007} suggested that a precessing warped Be disk caused the change of the profile from a normal double-peaked profile to a single-peaked one or a shell-like profile around the giant outbursts.
The detailed structure of the Be disk, however, remained unknown because observations have not been carried out intensively enough to cover those giant outburst.

Recently, A~$0535+262$/V725 Tau, which is one of the best studied Be/X-ray binaries since its discovery \citep{Rosenberg1975, Coe75}, was active in X-rays from 2008 to 2011.
This system consists of an X-ray pulsar with the spin period of 103 sec orbiting around an O9.7IIIe star \citep{Giangrande1980} in a relatively wide ($P_{\rm orb}$ $\sim$ $110$ days) and eccentric ($e\; \sim$ $0.47$) orbit \citep{Finger1994}.

In this X-ray active period A~$0535+262$ exhibited two giant outbursts as well as several normal outbursts.
In particular, the normal outbursts since 2009 April had peak X-ray luminosities several times higher than those before 2009 January.
\citet{Reynolds2010} reported X-ray  spectra of A~$0535+262$ in the fading phase of the 2009 November/December giant outburst by $Chandra$ grating spectroscopy.
They detected a significant feature associated with high velocity outflow ($>\;1000\;\mathrm{km\;s^{-1}}$) likely via radiatively driven disk wind from the accretion disk.
\citet{Camero-Arranz2012} summarized X-ray and optical data over thirty years, and found an anti-correlation between the equivalent width of the H$\alpha$ line and the $V$ magnitude around the giant outburst.
This anti-correlation was also discussed in \citet{Yan2012}, where the quantities from 1992 to 2010 were investigated.
They interpreted the decrease of the optical brightness during the giant outbursts as a result of mass ejection, which forms a tenuous region in the inner part of the Be disk.
\citet{Camero-Arranz2012} also reported that the V/R ratio of H$\alpha$, the ratio of the violet peak to the red one, showed short-term variations with a period of $\sim$ 25 days in late 2010 and before the 2011 February giant outburst.

We reported an overview of the H$\alpha$ and He I $\lambda$5876 line profile variabilities during and after the 2009 giant outburst in \citet{Moritani2011}, where we suggested the presence of a dense gas stream around the periastron and a warped Be disk.
In this paper, we carry out a full analysis of  all obtained data around the 2009 giant outburst as well as the data obtained after 2010 October, and discuss the Be disk structure in the recent X-ray active state.
In Section \ref{sec:obs} the configuration of our observations is summarized, and the results is described in section \ref{sec:result}.
The Be disk structure indicated from the observations is discussed in section \ref{sec:discuss}.
The conclusion is given in section \ref{sec:conclude}.

\section{Observations}\label{sec:obs}
\subsection{Spectroscopic Observations}
Optical spectroscopic observations of A~$0535+262$ were carried out on 75 nights from 2009 April to 2012 April at the Okayama Astrophysical Observatory (OAO) with a 188 cm telescope equipped with HIDES (High Dispersion Echelle Spectrograph), and at Gunma Astronomical Observatory (GAO) with a 1.5 m telescope equipped with GAOES (Gunma Astronomical Observatory Echelle Spectrograph).
Since 2009 December, most of the HIDES spectra were obtained with its new fiber-feed system \citep{Kambe2013}.
HIDES covers 3500 -- 6800 \AA \ range with three 2k $\times$ 4k e2V 42 -- 80 CCDs.
The wavelength coverage of GAOES is 4800 -- 6700 \AA \ and the detector is an e2V 44 -- 82 CCD.
The typical wavelength resolution $R$ of our HIDES and GAOES data around H$\alpha$ is $\sim$ 60000 and $\sim$ 30000, respectively.
Most of the H$\alpha$ data before 2010 March have signal to noise ratio $S/N$ of $\gtrsim$ 100, while the typical $S/N$ of the H$\alpha$ data after 2010 August is $\sim$ 30.
The effective exposure time was 1800 to 5400 seconds with HIDES, and 1800 to 9600 seconds with GAOES, respectively.

The obtained data were reduced in the standard way, using IRAF\footnote{http://iraf.noao.edu/} echelle package --- subtraction of bias, flat fielding, calibration of the wavelength using Th-Ar lines, normalization of the continuum, and helio-centric correction of the radial velocity.

\begin{figure*}
\begin{center}
	\FigureFile(150mm,100mm){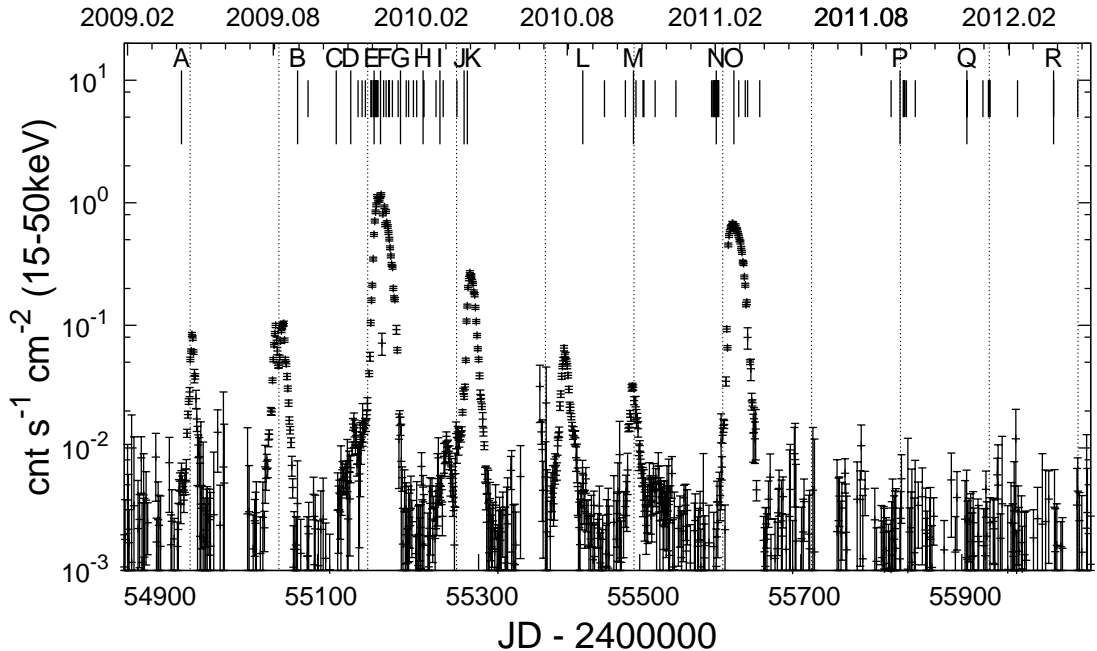}
\end{center}
\caption{X-ray (15 -- 50 keV) light curve by $Swift$/BAT archive data.
The vertical dotted lines indicate the rising time of the normal outburst [$\phi_X$ = 0 in \citet{Moritani2010}].
The lines above X-ray light curve correspond to the HJD of spectroscopic observations.
Among them, the long lines indicate HJD of the representative spectra shown in figures \ref{fig:profile_rep}, \ref{fig:profile_rep2} and \ref{fig:profile_rep3}.
}
  \label{fig:LC_X}
\end{figure*}

Figure \ref{fig:LC_X} shows the X-ray light curve of $Swift$/BAT\footnote{http://swift.gsfc.nasa.gov/docs/swift/results/transients/1A0535p262/} (15 -- 50 keV) for comparison with our observations.
The vertical dotted lines indicate the rising time of the normal outburst, $\phi_X$ = 0, estimated in \citet{Moritani2010}.
Our observations, indicated by the vertical lines above the X-ray light curve, cover the giant outburst in 2009 November/December, in the rising phase more densely than in the fading phase.
After the giant outburst, several monitoring observations were carried out until the next periastron passage in 2010 March, when a normal outburst occurred.
About half a year later, we started monitoring again, covering the giant outburst in 2011 February.
After the 2011 giant outburst, we carried out observations around a few more periastron passages, although A $0535+262$ stayed in an X-ray quiescent state.

Among the observed data, the representative H$\alpha$ and He I $\lambda$5876 line profiles before 2010 September are reported in \citet{Moritani2011}.

\subsection{Photometric Observations}
$B$ and  $V$-band photometry was performed from 2007 October to 2009 January at Osaka Kyoiku University Astronomical Observatory with a 51 cm telescope, along with the normal outburst in 2009 January [periods III and VIII of \citet{Moritani2010}].
The used CCD cameras are Astromed e2V 88200 until 2007 December, SBIG ST-9 from 2008 January to March, and Andor DW436 since 2008 April.

Log of photometric observations is listed in table \ref{tbl:log_p}.
Columns 1 and 2 list observation dates and mid exposure time in HJD.
$V$ magnitude and $B - V$ color are listed in columns 3 and 4, respectively.
On 2008 February 23, the error of $B-V$ color was larger than typical value since the cloudy condition made the $B$ magnitude poorly accurate. 
On 2008 September 10, we could not observe in $B$-band because of a bad weather.

\begin{table}[!t]
\caption{Photometric observation log.
Column 1 shows observational dates and column 2 shows HJD of each day.
$V$ magnitudes and $B - V$ colors were listed in columns 3 and 4, respectively.
}
\label{tbl:log_p}
\begin{center}
\begin{tabular}{cccc}\hline \hline
Date	& HJD	& $V$[mag]	& $B - V$ [mag]	\\ \hline
2007.10.20	& 2454394.25	& 8.83$\pm$0.01	& 0.64$\pm$0.01	\\
2007.10.21	& 2454395.13	& 8.85$\pm$0.01	& 0.62$\pm$0.02	\\
2007.10.31	& 2454405.12	& 8.89$\pm$0.01	& 0.61$\pm$0.01	\\
2007.11.07	& 2454412.13	& 8.91$\pm$0.02	& 0.58$\pm$0.02	\\
2007.11.14	& 2454419.18	& 8.90$\pm$0.01	& 0.56$\pm$0.02	\\
2007.11.24	& 2454429.08	& 8.92$\pm$0.02	& 0.58$\pm$0.03	\\
2007.11.30	& 2454435.10	& 8.88$\pm$0.02	& 0.60$\pm$0.02	\\
2007.12.04	& 2454439.12	& 8.86$\pm$0.01	& 0.62$\pm$0.01	\\
2007.12.19	& 2454453.99	& 8.86$\pm$0.01	& 0.60$\pm$0.01	\\
2007.12.26	& 2454461.16	& 8.88$\pm$0.01	& 0.55$\pm$0.02	\\
2008.01.24	& 2454489.93	& 8.94$\pm$0.01	& 0.55$\pm$0.01	\\
2008.02.23	& 2454520.18	& 8.93$\pm$0.03	& 0.54$\pm$0.06	\\
2008.03.10	& 2454536.13	& 8.94$\pm$0.02	& 0.53$\pm$0.02	\\
2008.03.17	& 2454543.04	& 8.95$\pm$0.01	& 0.56$\pm$0.01	\\
2008.07.26	& 2454674.28	& 8.89$\pm$0.03	& 0.60$\pm$0.04 	\\
2008.08.24	& 2454703.27	& 8.87$\pm$0.01	& 0.61$\pm$0.01	\\
2008.09.10	& 2454720.21	& 8.86$\pm$0.02	& --	\\
2008.10.15	& 2454755.27	& 8.89$\pm$0.01	& 0.59$\pm$0.01	\\
2008.10.16	& 2454756.14	& 8.87$\pm$0.01	& 0.60$\pm$0.01	\\
2008.11.12	& 2454783.26	& 8.92$\pm$0.01	& 0.59$\pm$0.01	\\
2008.11.13	& 2454784.10	& 8.91$\pm$0.01	& 0.59$\pm$0.01	\\
2008.11.18	& 2454789.26	& 8.91$\pm$0.01	& 0.59$\pm$0.01	\\
2008.11.19	& 2454790.33	& 8.91$\pm$0.02	& 0.59$\pm$0.02	\\
2008.11.26	& 2454797.32	& 8.90$\pm$0.02	& 0.59$\pm$0.02	\\
2008.12.01	& 2454802.00	& 8.90$\pm$0.01	& 0.58$\pm$0.01	\\
2008.12.02	& 2454802.97	& 8.89$\pm$0.01	& 0.60$\pm$0.01	\\
2008.12.03	& 2454804.30	& 8.88$\pm$0.01	& 0.59$\pm$0.01	\\
2008.12.06	& 2454807.06	& 8.90$\pm$0.01	& 0.59$\pm$0.01	\\
2008.12.10	& 2454811.28	& 8.90$\pm$0.01	& 0.60$\pm$0.01	\\
2008.12.21	& 2454824.20	& 8.89$\pm$0.01	& 0.58$\pm$0.01	\\
2008.12.23	& 2454838.17	& 8.88$\pm$0.01	& 0.60$\pm$0.01	\\
2009.01.13	& 2454845.10	& 8.90$\pm$0.01	& 0.59$\pm$0.01	\\
2009.01.15	& 2454847.13	& 8.90$\pm$0.01	& 0.59$\pm$0.01	\\
2009.01.19	& 2454851.22	& 8.88$\pm$0.01	& 0.60$\pm$0.01	\\
2009.01.20	& 2454852.09	& 8.88$\pm$0.01	& 0.60$\pm$0.01	\\ \hline
\end{tabular}    
\end{center}
\end{table}

\section{Results}\label{sec:result}
Observed line profiles exhibit remarkable variabilities not only in the global shape but also in many characterizing quantities.
In this section, we describe observed variabilities in detail.
We summarize principal features in subsection \ref{subsec:sum_obs} in order to discuss the structure of the Be disk.

\subsection{Profile Variabilities}
\begin{figure*}[!p]
\begin{center}
\begin{tabular}{c}
		\FigureFile(160mm,90mm){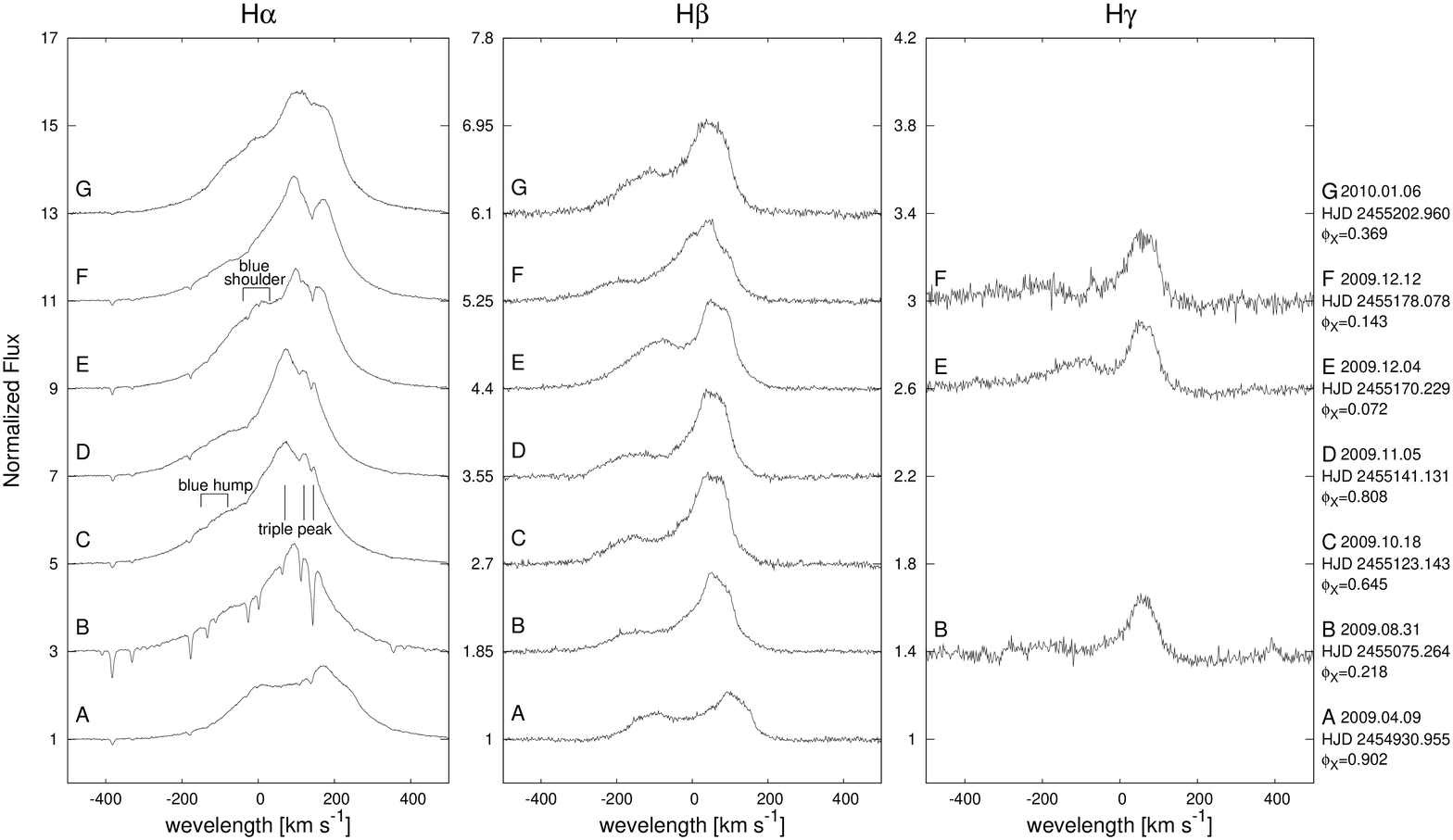} \\
		\FigureFile(160mm,90mm){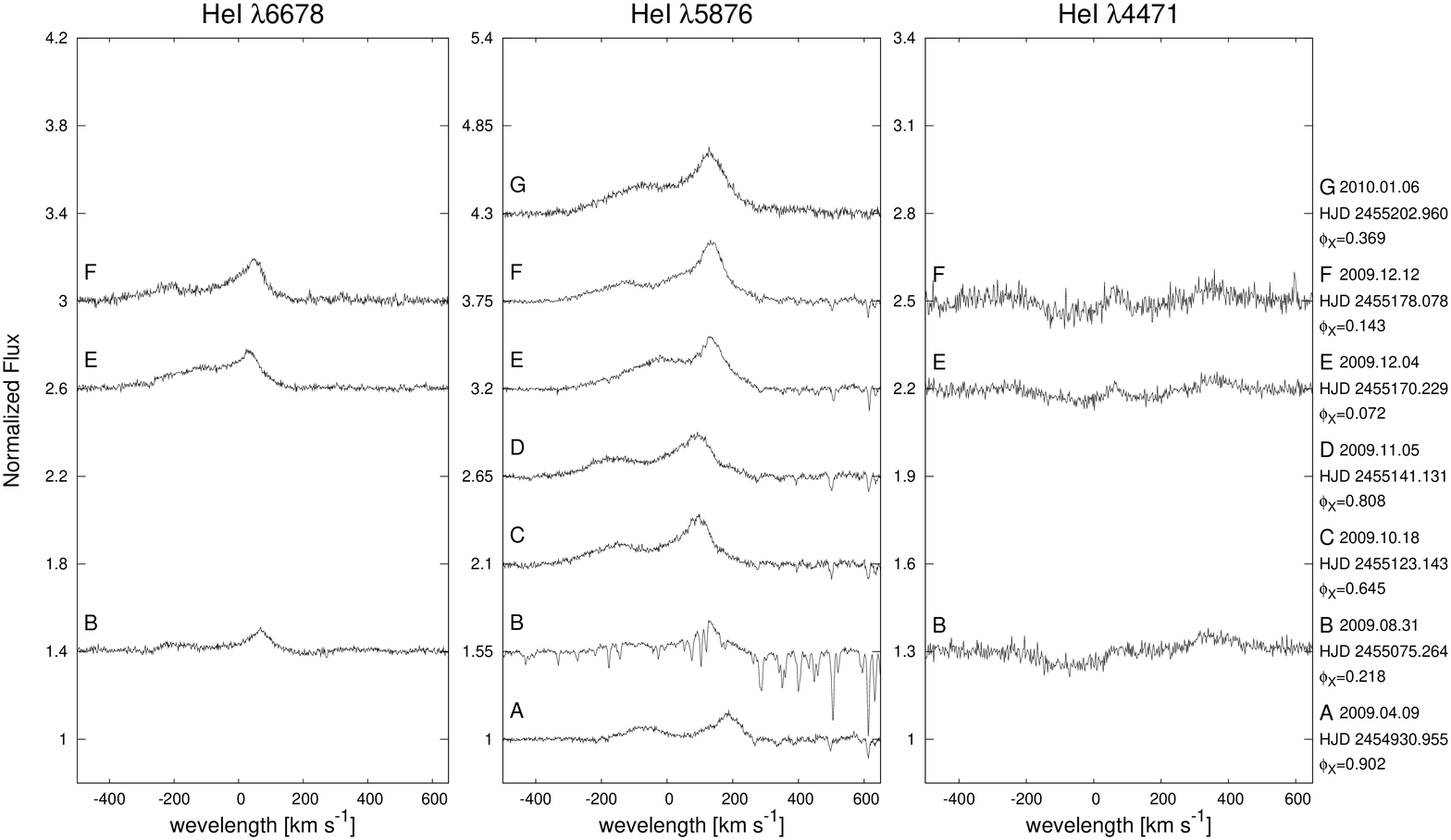} \\
\end{tabular}
\end{center}
\caption{
Representative spectra before and during the giant outburst in 2009.
In the upper panel, the H$\alpha$, the H$\beta$ and the H$\gamma$ line profiles are shown from the left, while in the lower panel, the He I $\lambda$6678, the He I $\lambda$5876 and the He I $\lambda$4471 are displayed from the left. 
For the sake of clarity, the profiles are shifted from each other in the vertical direction.
The date, HJD and $\phi_X$ of each spectrum are annotated on the right side of the figure.
Note that the horizontal scale is different between the Balmer and the He I line profiles.
}
\label{fig:profile_rep}
\end{figure*}

The representative H$\alpha$, H$\beta$, H$\gamma$, He I $\lambda$6678, He I $\lambda$5876 and He I $\lambda$4471 line  profiles are shown in figures \ref{fig:profile_rep} -- \ref{fig:profile_rep3}, which display the profiles observed before and during the giant outburst in 2009 (subsection \ref{subsec:lpv2009}), between the 2009 and 2011 giant outbursts (subsection \ref{subsec:lpv2009_2011}) and during and after the giant outburst in 2011 (subsection \ref{subsec:lpv2011}), respectively.
For the sake of clarity, the spectra have linear offsets along the vertical axis from each other. 
The observation date, HJD of the mid-exposure time and  the orbital phase $\phi_X$ are annotated on the right side of panels.
All observed profiles are shown in the appendix (figures \ref{fig:profile_all1} and \ref{fig:profile_all2}).

Figures \ref{fig:profile_rep} -- \ref{fig:profile_rep3} show that the H$\alpha$ and the H$\beta$ line profiles drastically changed in the following observational periods: before the giant outburst in 2009 (spectra A -- D), during it (spectra E -- G), between the giant outbursts in 2009 and in 2011 (spectra H -- N), and afterward (spectra O -- R). 
The H$\alpha$ line profiles from 2009 August to 2010 March were characterized by a strongly redshifted triple peak and a broad hump/shoulder in the blue wing.
The H$\beta$ line profiles in that period also had a strongly redshifted multi-peaked component.
After the normal outburst in 2010 March, the H$\alpha$ and the H$\beta$ line profiles gradually weakened with changing V/R ratios.
The H$\gamma$ line, which was observed only during the giant outburst in 2009, showed a complicated profile with both the emission component and the absorption component.
Its emission component appeared to exhibit variability in phase with the H$\alpha$ and the H$\beta$ lines.

The variation of the He I $\lambda$6678 and the He I $\lambda$5876 line profiles was also significant, although it was not as drastic as that of the H$\alpha$ or the H$\beta$ line.
Moreover, they exhibited not only typical double-peaked profiles observed in many Be stars (until around the normal outburst in 2010 March, spectra A -- J), but also complicated triple-peaked profiles around the normal outburst in 2010 October and the 2011 giant outburst (spectra K -- N).
The He I $\lambda$4471 line profile showed a broad absorption feature containing an emission component at the center.
The emission component of the He I $\lambda$4471 line profiles also changed in both its peak flux and position.

In the following sections, the variabilities in each period are described.

\subsubsection{Before and during the Giant Outburst in 2009}\label{subsec:lpv2009}

In 2009 April (Spectrum A), around one cycle before the ``double-peaked" normal outburst \citep{Caballero2009}, the H$\alpha$, the H$\beta$ and the He I $\lambda$5876 lines exhibited double-peaked profiles with the red peak higher than the violet peak (V $<$ R).

After the ``double-peaked" normal outburst in 2009 August (Spectrum B), the H$\alpha$ line exhibited almost single-peaked profile with a bright red component (V $\ll$ R), whereas the H$\beta$ and the He I $\lambda$5876 line profiles were double-peaked with V $\ll$ R.
Although the H$\gamma$ and the He I $\lambda$6678 lines had similar double-peaked profiles with the red peak brighter than the violet peak in 2009 August, an absorption wing was seen in the red side of the H$\gamma$ line profile.
A double-peaked emission component was also seen in the red side of the He I $\lambda$4471 absorption line profile (between $\sim\;+100\;\mathrm{km\;s^{-1}}$ and $\sim\;+400\;\mathrm{km\;s^{-1}}$).

In one month, during the precursors of the 2009 giant outburst in 2009 October and November \citep{Wilson2009,Sugizaki2009}, the flux of the continuum-normalized peak of the H$\alpha$, the H$\beta$ and the He I$\lambda$5876 lines significantly increased, and the blue hump superposed on the smooth slope of the H$\alpha$ line brightened (spectra C and D).
The blue component of the H$\beta$ and the He I $\lambda$5876 line profiles also became brighter.

In the rising phase of the giant outburst in 2009, the H$\alpha$ line profile changed drastically: the peak flux decreased and a bright ``shoulder" appeared around 0 $\mathrm{km\;s^{-1}}$ in the blue wing (spectrum E), as reported in \citet{Moritani2011}.
Hereafter we call this bright feature blue shoulder.
A similar bright feature was observed in the central part of the H$\beta$, the He I $\lambda$6678 and the He I $\lambda$5876 line profiles (between $-50\;\mathrm{km\;s^{-1}}$ and $+50\;\mathrm{km\;s^{-1}}$).
However, these features were gone by the peak of the giant outburst --- 2009 December 10 \citep{Krimm2009,Caballero2010}.
In the H$\gamma$ line, the blue component brightened once and weakened again, and the central dip shifted blue-ward (from $\sim\;0\;\mathrm{km\;s^{-1}}$ to $\sim\; -100\;\mathrm{km\;s^{-1}}$).
The emission component of the He I $\lambda$4471 line profile has shifted blue-ward from 2009 August to December, but seems not to have changed during the X-ray brightening period.

Two days after the peak of the giant outburst, 2009 December 12, the H$\alpha$ and the H$\beta$ line profile had double peak in their enhanced red side (spectrum F).
The H$\gamma$ line seemed to have a small peak in the center ($\sim\; -100\;\mathrm{km\;s^{-1}}$), in addition to the double-peaked profile.

Three weeks later (6 January 2010), in the fading phase of the giant outburst, the H$\alpha$ line profile exhibited a quadruple peak --- one double peak in the blue side and the other double peak in the red side (spectrum G).
The double peak in the red side of the H$\beta$ line merged into a single peak. 
The peak flux of these lines had started decreasing by then, after its gradual increase until 2009 December 22.
The He I $\lambda$5876 line showed similar variability: the peak flux increased and then slightly decreased again.

\subsubsection{After the Giant Outburst in 2009 and Before the Giant Outburst in 2011}\label{subsec:lpv2009_2011}

\begin{figure*}[!p]
\begin{center}
\begin{tabular}{c}
		\FigureFile(160mm,90mm){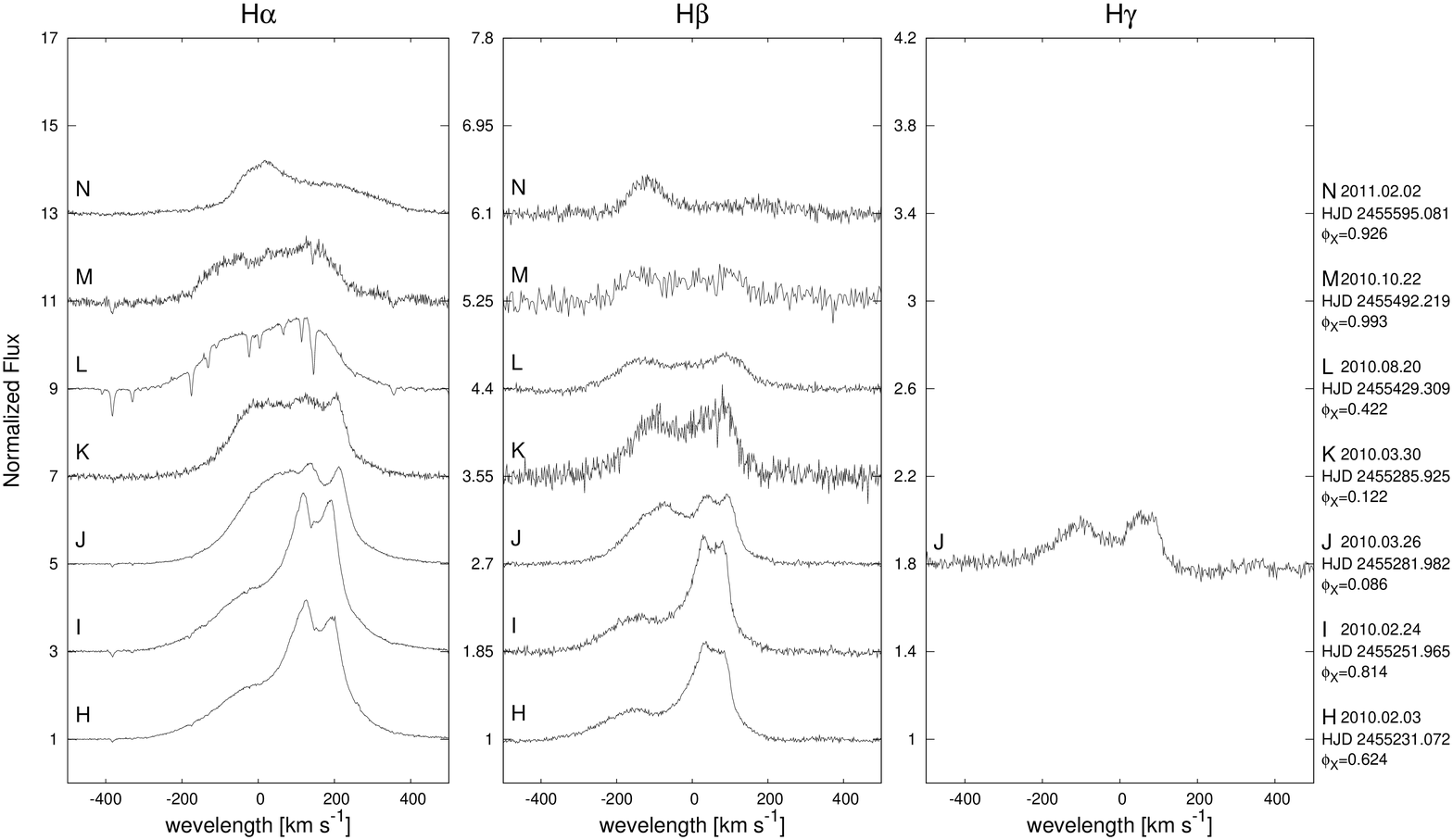} \\
		\FigureFile(160mm,90mm){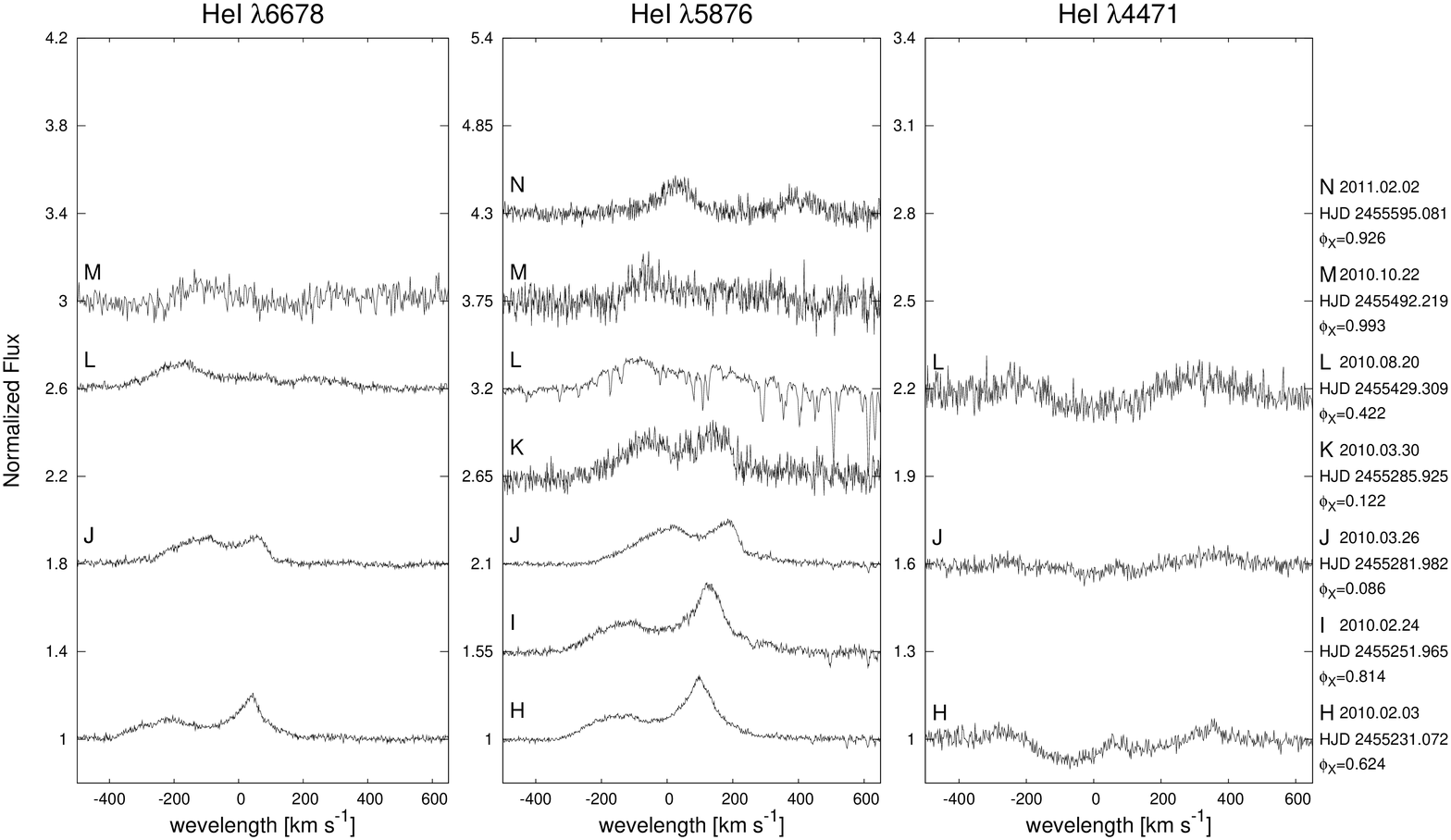} \\
\end{tabular}
\end{center}
\caption{
Same as figure \ref{fig:profile_rep}, but for representative spectra between the giant outburst in 2009 and in 2011.
}
\label{fig:profile_rep2}
\end{figure*}

In 2010 February, the redshifted double-peaked component in the H$\alpha$ and the H$\beta$ lines continued to grow (spectra H and I), and became the highest during the precursor of the next normal outburst (2010 February 24, spectrum I).
The peak flux of the He I $\lambda$5876 also increased.
The violet component of the He I $\lambda$6678 became brighter and broader than that during the 2009 giant outburst.
On the contrary to other lines, the peak flux of the emission component of the He I $\lambda$4471 weakened compared to that in the X-ray maximum phase of the 2009 giant outburst. 

On 2010 March 26, near the peak of the normal outburst in the next orbital cycle of the giant outburst, the profile drastically changed (spectrum J): in the H$\alpha$ line, the strongly redshifted double-peaked component weakened and a huge, broad shoulder appeared in the blue wing.
In the H$\beta$ line the redshifted double-peaked component also weakened, while the violet peak brightened and shifted red-ward (from $\sim\;-150\;\mathrm{km\;s^{-1}}$ to $\sim\;-100\;\mathrm{km\;s^{-1}}$).
Other emission lines showed a similar trend to the H$\alpha$ and the H$\beta$ lines.
The red side of the H$\gamma$ line profile became weak and double-peaked, while the blue side became as bright as the red side (V $\sim$ R).
In the He I $\lambda$6678 and the He I $\lambda$5876 line profiles, a bright component appeared with a similar velocity range to that of the blue hump in the H$\beta$ and the H$\alpha$ lines, respectively.
The emission component of the He I $\lambda$4471 weakened and the absorption component in the blue side disappeared.

In the next four days, the bright blue shoulder grew wider and the H$\alpha$ line profile became like a top-hat profile (spectrum K).
The H$\beta$ line profile, on the contrary, did not seem to have changed, although relatively poor $S/N$ of this profile makes it difficult to compare them in detail.
The He I $\lambda$5876 line profile shifted blue-ward.

By 2010 August 20, or two orbital cycles after the giant outburst in 2009, all observed lines had significantly weakened (spectrum L).
Double- or triple-peaked component disappeared in the red side of the H$\alpha$ line profile.
The H$\beta$ line exhibited a relatively normal double-peaked profile as for a Keplerian disk except that a small peak was seen at the center ($\sim \; -50\; \mathrm{km\;s^{-1}}$).
This feature disappeared by 2010 October 12, two weeks before the third normal outburst from the giant outburst in 2009.
The V/R ratios of the He I $\lambda$6678 and the He I $\lambda$5876 line profiles turned to $<$ 1, while those of the H$\alpha$ and the H$\beta$ line profiles remained $>$ 1.
The He I $\lambda$4471 line profile became rather symmetric with the emission wings and the central absorption.

At the next periastron, during the normal outburst in 2010 October (spectrum M), the H$\alpha$ line profile slightly changed to have a small dip and a small peak around $-20\;\mathrm{km\;s^{-1}}$ and $+20\;\mathrm{km\;s^{-1}}$ respectively.
The separation of the double peak of the H$\beta$ profile became narrower, and the peak flux of the violet and red peaks was almost the same (V $\sim$ R). 
After the normal outburst, the V/R ratio of the H$\alpha$ turned to $>$ 1 (JD $\sim$ 2455500, see also figure \ref{fig:X_VoverR}) and exhibited a variation in the range of $\gtrsim$ 1.
\citet{Camero-Arranz2012} reported that the H$\alpha$ exhibited short-term ($\sim$ 25 days) V/R variation in this period (see their figure 4).
The H$\beta$, however, seemed to continue to increase its V/R ratio.
The He I $\lambda$6678 and the He I $\lambda$5876 lines exhibited triple-peaked profile in this period, while the He I $\lambda$4471 did not changed, compared to that two months before.

One week before the giant outburst in 2011 (spectrum N), the H$\alpha$ line  profile only had a wide red wing without enhanced component or triple-peaked feature as shown in the previous giant outburst.
The blue hump appeared in the wing ($\sim \; -50\;\mathrm{km\;s^{-1}}$).
The H$\beta$ line also exhibited V $\gg$ R profile, whose red peak was so broad that it was barely seen.
The He I$\lambda$5876 line showed a double-peaked profile with V $>$ R.

\subsubsection{During and After the Giant Outburst in 2011}\label{subsec:lpv2011}

\begin{figure*}[!p]
\begin{center}
\begin{tabular}{c}
		\FigureFile(160mm,90mm){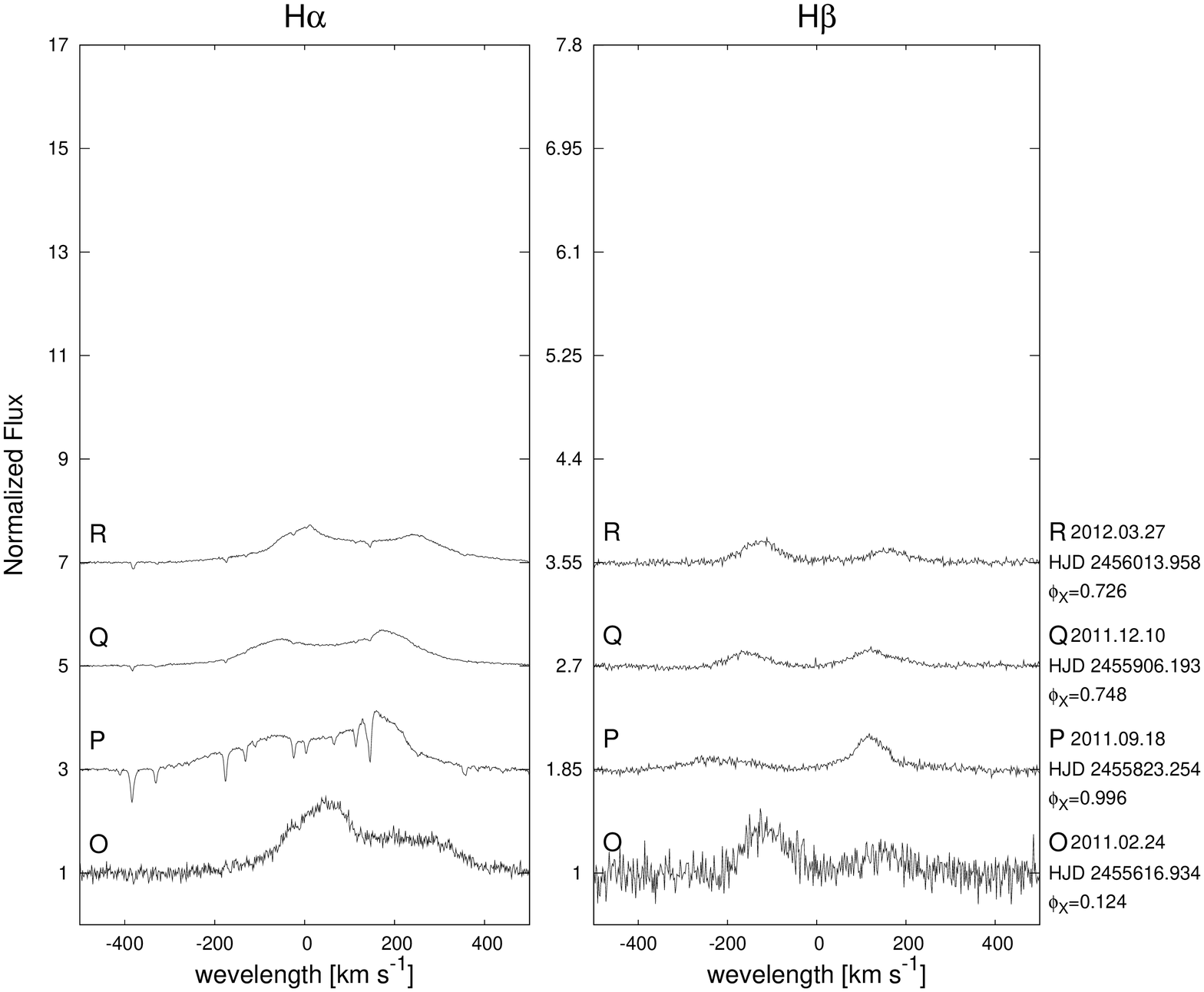} \\
		\FigureFile(160mm,90mm){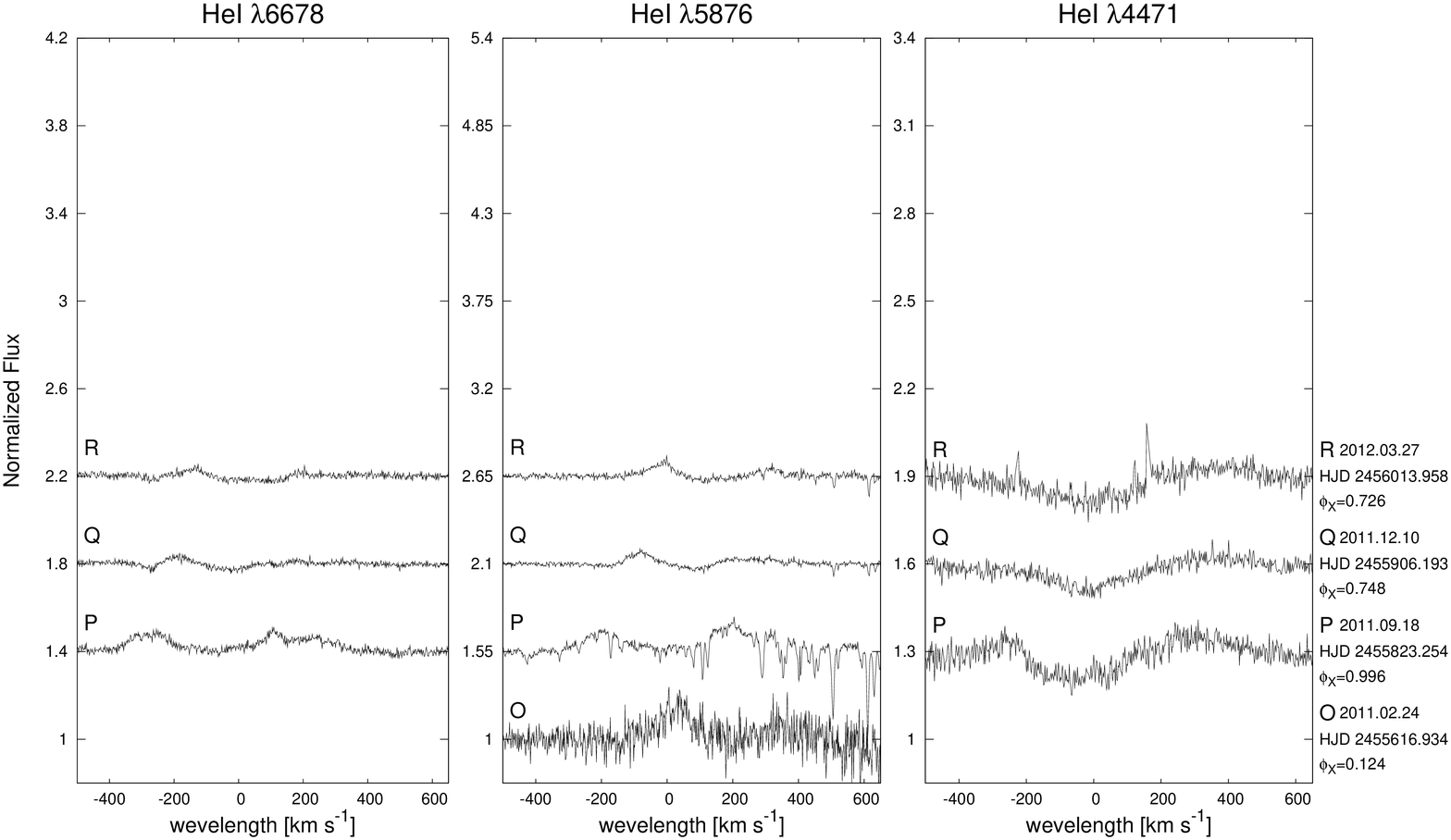} \\
\end{tabular}
\end{center}
\caption{
Same as figure \ref{fig:profile_rep}, but for representative spectra during and after the giant outburst in 2011.
}
 \label{fig:profile_rep3}
\end{figure*}

Around the X-ray maximum flux of the 2011 giant outburst (2011 February 24, spectrum O), a hump showed up in the red wing of the H$\alpha$ line profile ($\sim \; 300 \;\mathrm{km\;s^{-1}}$) and the red side of the profile became rather flat.
In a month, the center of the H$\alpha$ line profile brightened once and weakened again, and a flat shape re-appeared on 2011 March 28. 
The red side of the H$\beta$ line profile became slightly brighter around the X-ray maximum, and the center dip became deeper and narrower by the end of March.
The He I $\lambda$5876 line profile did not vary during the 2011 giant outburst, although $S/N$ ratios of the profiles taken in this period were not good enough to investigate variabilities in detail.

Around the periastron two cycle after the giant outburst in 2011 (2011 September 18), the H$\alpha$ and the H$\beta$ lines exhibited  V $<$ R profiles (spectrum P).
In particular, the blue side of the H$\beta$ line was broad.
The V/R ratios of the He I $\lambda$6678 and the He I $\lambda$5876 were also $<$ 1, although that of  the He I $\lambda$6678 was $\sim$ 1.
The red side of the He I $\lambda$6678 line profile had a double peak.
There was a hint of the similar feature in the He I $\lambda$5876 line, but it was difficult to clearly see it because of the atmospheric absorption lines.
The He I $\lambda$4471 line profile was similar to that in 2011 March, but the center of the absorption shifted blue-ward.
No outburst was detected in X-rays at this periastron passage.
Neither was the line profile variability.

Recently, A $0525+262$ stayed in a quiescent state in X-rays.
Observed H$\alpha$, H$\beta$, He I $\lambda$6678 and He I $\lambda$5876 lines remain in emission, although it slightly weakened (Spectra Q and R).
The He I $\lambda$4471 line turned to an almost absorption profile with weak broad emission component in the red wing.
The V/R ratios of these lines varied, almost in phase.

\subsection{Other Variabilities}
The equivalent width, the peak flux and the V/R ratio of each line profile also showed various variabilities.
In this section, the variabilities of these quantities are described.
The obtained equivalent widths and V/R ratios of each line are summarized in table \ref{tbl:Obs_results} (see appendix \ref{app:obs}).

\subsubsection{Equivalent Widths and Peak Fluxes}

\begin{figure*}[p]
 \begin{center}
  \FigureFile(145mm,250mm){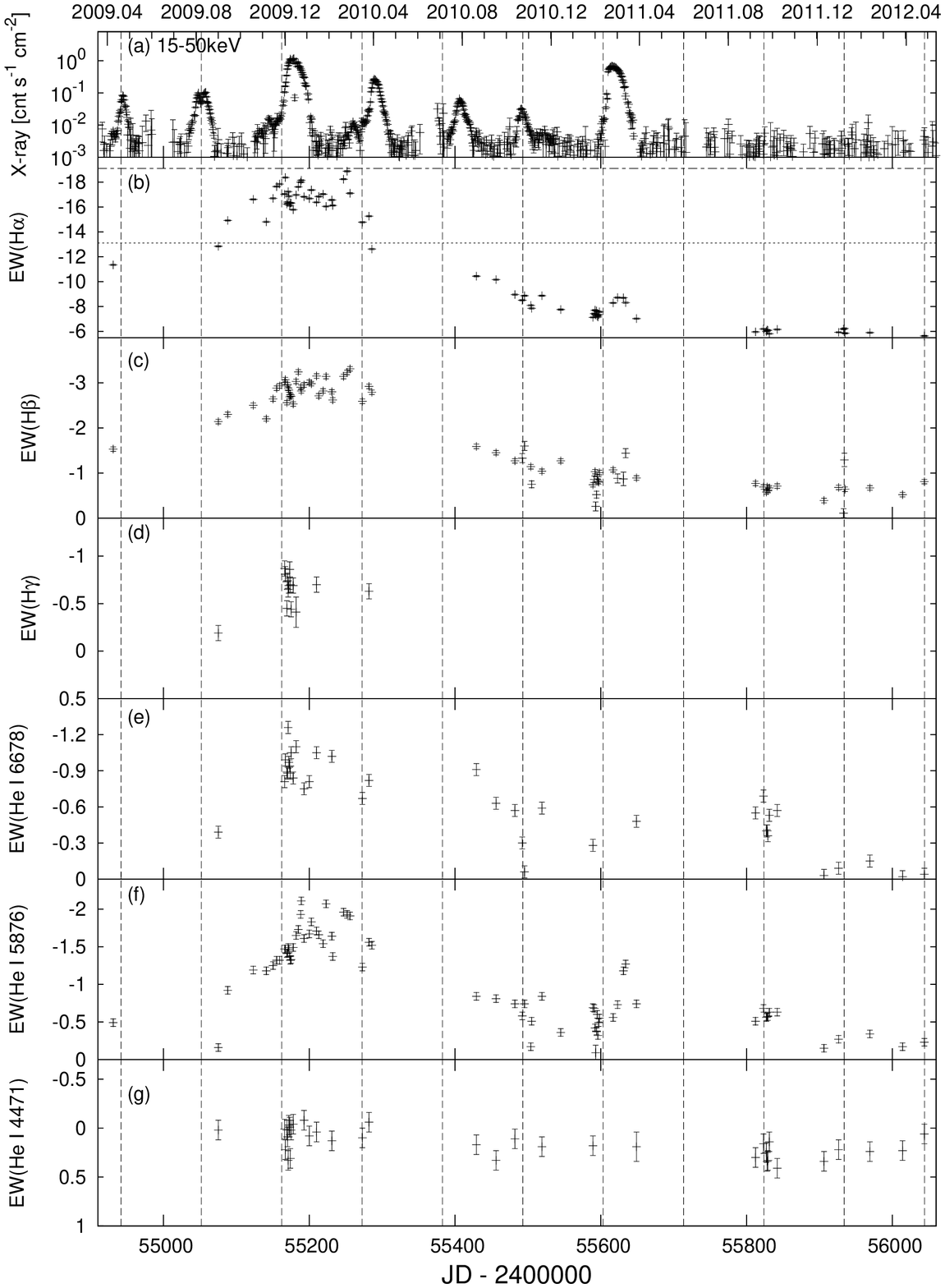}
 \end{center}
 \caption{
Equivalent widths of  H$\alpha$ (b), H$\beta$ (c), H$\gamma$ (d), He I $\lambda$6678 (e), He I $\lambda$5876 (f) and He I $\lambda$4471 (g) lines from 2009 April to 2011 April.
For comparison with X-ray activity, the light curve by $Swift$/BAT are shown in panel (a).
Vertical dashed lines indicate the time at $\phi_X = 0$.
Horizontal dotted and dashed-dotted lines in panel (b) denote the expected values for the Be disk filling the Roche lobe radius at periastron (5.6 $R_*$) in the case of $V$ = 8.9 ($-13$ \AA ) and $V$ = 9.3 ($-19$ \AA ), respectively (see section \ref{sec:warp}).
}
 \label{fig:X_EW}
\end{figure*}

In the last seven years, the equivalent width of each observed line (figure \ref{fig:X_EW}) was the highest in its absolute value (i.e. the emission line was the strongest) around the giant outburst in 2009  --- $EW\mathrm{(H\alpha)} \; \sim \; -18$ \AA ,  $EW\mathrm{(H\beta)} \; \sim \; -3$ \AA , $EW\mathrm{(He I\;\lambda6678)} \; \sim \; -0.9$ \AA , $EW\mathrm{(He I\;\lambda5876)} \; \sim \; -1.5$ \AA .
$EW\mathrm{(He I\;\lambda4471)}$ was almost constant at $\sim \; 0$ \AA, since the line exhibited a blended profile with absorption and emission components, but slightly smaller during the giant outburst in 2009, when the line profile had stronger emission.

$EW\mathrm{(H\alpha)}$ showed complicated variability during the monitoring period.
At the beginning of the observed period, $|EW\mathrm{(H\alpha)}|$ [the absolute value of $EW\mathrm{(H\alpha)}$] monotonically increased at least for one orbital period, except for a temporary decrease at the X-ray precursor in 2009 October ($\sim$ JD 2455140) 
Afterwards, it increased for a while until ten days before the peak of the 2009 giant outburst ($\sim$ JD 2455170).
A brief low state in the H$\alpha$ line flux was seen for $\sim$ 10 days between JD 2455169 and JD 2455178 (2009 December 3 -- 12), which was followed by a gradual decrease of $|EW\mathrm{(H\alpha)}|$ until the end of the giant outburst.
Then, after a rapid increase two weeks prior to the precursor of the normal outburst in 2010 March ($\sim$ JD 2455270), it entered a long-term, declining phase, which continued until 2011 February ($\sim$ JD 2455600), with a short-term (less than 10 days) increase during three successive normal outbursts.
\citet{Camero-Arranz2012} and \citet{Yan2012} reported a similar trend of variability in $|EW\mathrm{(H\alpha)}|$, although $|EW\mathrm{(H\alpha)}|$ values of this work are approximately 5 \AA \ smaller than theirs\footnote{
The spectral resolution of their H$\alpha$ data ($R\;\sim$ 5000) was lower than that of this work ($R\;\sim$ 30000 -- 60000).
This made the H$\alpha$ line broader.
Besides, their spectra had wider wings in the red side (up to $\sim \; \pm$ 1000 $\mathrm{km\;s^{-1}}$).
Therefore, there is a difference in drawn continuum between their observations and ours. 
(Camero-Arranz and Yan, in private communication)
}.
During the period of the giant outburst in 2011 ($\sim$ JD 2455630), $|EW\mathrm{(H\alpha)}|$ increased by 1 \AA ($\sim \;-8$ \AA \ to $\sim \;-9$ \AA ), and then returned to the previous level before at the end of the outburst ($\sim$ JD 2455650).
After the 2011 giant outburst, $|EW\mathrm{(H\alpha)}|$ decreased down to $\sim \; 6$ \AA , which was lowest in the last seven years.
$|EW\mathrm{(H\beta)}|$, $|EW\mathrm{(He I\;\lambda6678)}|$  and $|EW\mathrm{(He I\;\lambda5876)}|$ showed similar variability to that of$|EW\mathrm{(H\alpha)}|$.

The peak flux of each line showed a similar trend to that of the absolute value of the equivalent width, because the strong emission was associated with the redshifted enhanced component (e.g. spectra B -- I).
The peak fluxes of the H$\alpha$ and the H$\beta$ lines declined rapidly when the redshifted bright component vanished during the  normal outburst in 2010 March (spectrum J).
Afterwards the peak flux of all lines gradually decreased, although it increased again around the 2011 giant outburst.

\subsubsection{V/R Ratio}\label{subsec:VoerR}

\begin{figure*}[tb]
 \begin{center}
  \FigureFile(160mm,80mm){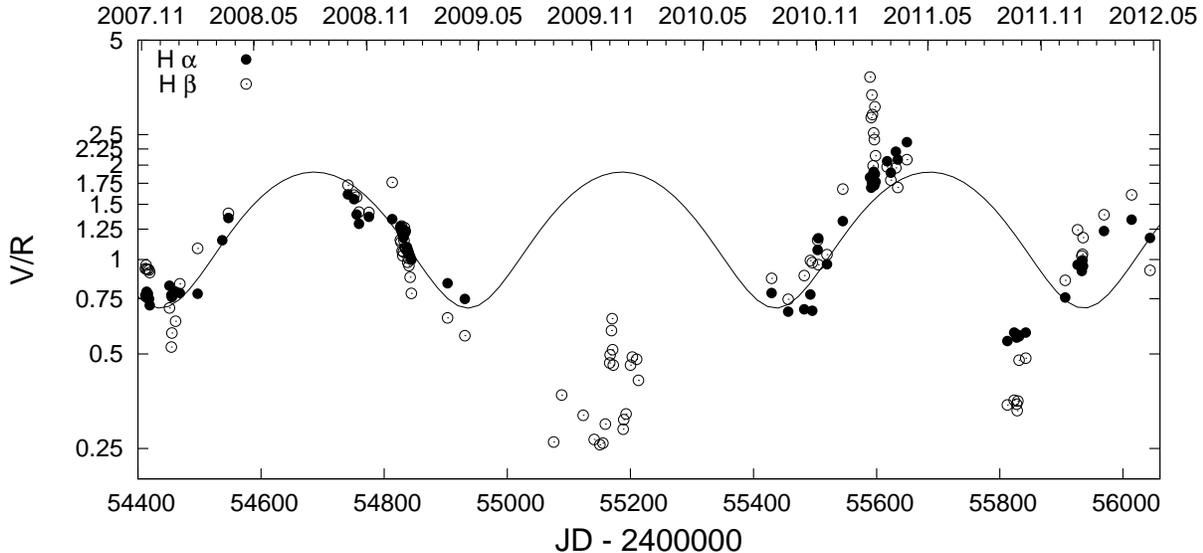}
 \end{center}
 \caption{
V/R ratios of the H$\alpha$ (filled circle) and the H$\beta$ (open circle) lines from 2007 November to 2011 October.
Sold line indicates the best-fit sine curve presented in \citet{Moritani2010}.
}
 \label{fig:X_VoverR}
\end{figure*}

Figure \ref{fig:X_VoverR} shows the V/R ratios of the H$\alpha$ (filled circle) and the H$\beta$ (open circle) lines from JD 2454400 to JD 2456060 (from 2007 November to 2012 April).
In this figure the data before 2009 March \citep{Moritani2010} are also plotted to check the tendency more easily.
The V/R ratio of the H$\alpha$ line was $\ll$ 1 around the giant outburst in 2009 ($\sim$ JD 2455050 -- JD 2455300), but the ratio itself cannot be determined because the redshifted enhanced component had more than one peak (labeled with ``n/a" in table \ref{tbl:Obs_results}).
The solid line in the figure is a best-fit sine curve for the H$\alpha$ line presented in \citet{Moritani2010}.

After the ``double-peaked" normal outburst in 2009 August, the V/R ratio unexpectedly became $\ll$ 1, which was caused by the redshifted enhanced component, until around the normal outburst in 2010 October.
In this period the V/R ratios of other lines were also $\ll$ 1.

Recently, the 500-day periodic variations presented by \citet{Moritani2010} no more traces the observed V/R ratios, in particular the data after the giant outburst in 2011.
This is possibly because the giant outbursts changed the structure of the Be disk so much that the parameters of the density wave also had changed.
Note that no valid function is obtained for the V/R ratios for the period only after the giant outburst in 2011 ($>$ 2455700) because of the lack of data.

\subsubsection{Wing of the H$\alpha$ Line Profiles}
\begin{figure*}[th]
\begin{center}
\FigureFile(160mm,250mm){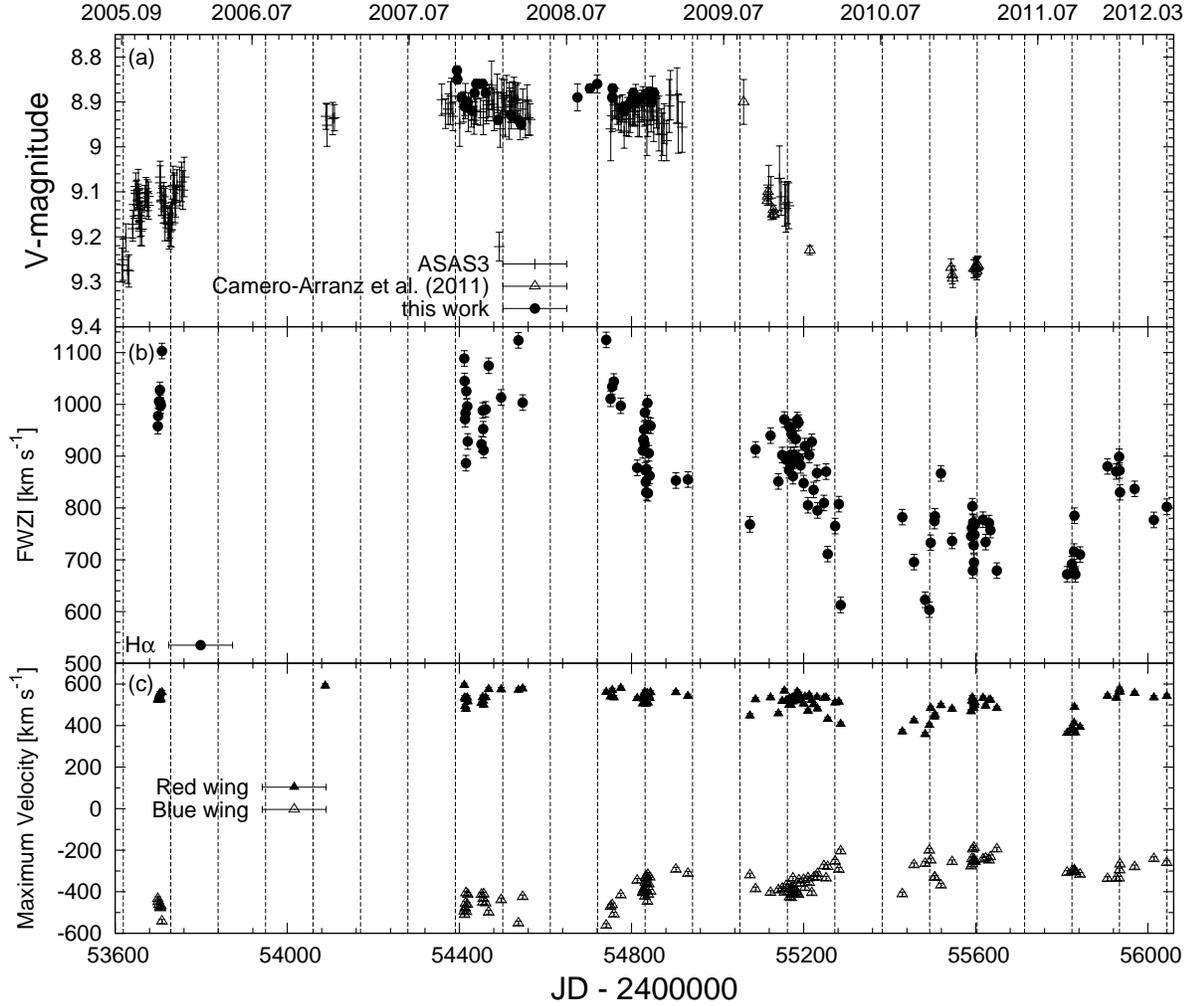}
\end{center}
\caption{
Wing parameters of the H$\alpha$ line profile and $V$ magnitude from 2005 September to 2012 April.
\textit{Top panel}): $V$ magnitude from ASAS3 (pluses), \citet{Camero-Arranz2012} (open triangles) and this work (filled circles).
\textit{Middle panel}: Full Width of Zero Intensity of the H$\alpha$ line.
\textit{Bottom panel}: Maximum velocity of the blue side (open triangle) and the red (filled triangle) side.
The vertical dashed lines indicate $\phi_X = 0$.
}
 \label{fig:Wing_Halpha}
\end{figure*}

Figure \ref{fig:Wing_Halpha} shows the variations of $V$ magnitude (top panel), the Full Width of Zero Intensity (FWZI, middle panel) and the maximum velocities (bottom panel) of the H$\alpha$ line from 2005 November to 2012 April (JD 2453500 -- JD 2456060). 
The vertical dashed lines indicate periastron passages ($\phi_X$ = 0).
$V$ magnitudes are taken from ASAS3\footnote{http://www.astrouw.edu.pl/asas/?page=main} (pluses), \citet{Camero-Arranz2012} (open triangles) in addition to this work (filled circles).
The maximum velocities of the H$\alpha$ line profiles, $RV_{\mathrm{wing,B}}$ and $RV_{\mathrm{wing,R}}$, are defined by the following steps because the profile was often asymmetric having a long tailed wing on one side.
Firstly, the intersections between $F_{\lambda}/F_{\lambda , \mathrm{continuum}} = 1.1$ and a Gaussian function that best fits the observed profile with $F_{\lambda}/F_{\lambda , \mathrm{continuum}} \leq 1.2$ are determined as $\lambda_1$ and $\lambda_2$ ($\lambda_1 < \lambda_2$). 
In the second step, the equivalent widths, $EW$(H$\alpha,_{V\leq1.1}$) and $EW$(H$\alpha,_{R\leq1.1}$), are estimated for the region outside of $\lambda_1$ and $\lambda_2$ ($\lambda < \lambda_1$ and $\lambda_2 < \lambda$, respectively).
Finally, the maximum velocities are estimated by shifting the intersections for $EW$(H$\alpha,_{\leq1.1}$).
That is,
\begin{equation}
RV_{\mathrm{wing,B}} = \lambda_1 - 20 \times |EW(\mathrm{H\alpha,_{V\leq1.1}})|
\end{equation}
for the violet wing, and
\begin{equation}
RV_{\mathrm{wing,R}} = \lambda_2 + 20 \times |EW(\mathrm{H\alpha,_{R\leq1.1}})|
\end{equation}
for the red wing.
The factor $20$ is needed in order to treat $EW$(H$\alpha,_{\leq1.1}$) in the same way as the wavelength, because we approximate the region outside of $\lambda_1$ and $\lambda_2$ by a triangle.
The obtained data are summarized in table \ref{tbl:Obs_wing} (appendix \ref{app:obs}).
FWZI is determined as difference between $RV_{\mathrm{wing,B}}$ and $RV_{\mathrm{wing,R}}$.

Top panel of figure \ref{fig:Wing_Halpha} shows that the optical brightness started to fade a couple of cycle before the giant outburst 2009 (JD $\sim$ 2455000), as \citet{Camero-Arranz2012} and \citet{Yan2012} reported.
The FWZI, on the other hand, seems to start decreasing a few cycle before the $V$ magnitude began fading, by the beginning of the 2008 (JD $\sim$ 2454500), although it occasionally increased around periastron.
After the 2011 giant outburst it slightly increased for one year.
The maximum velocities oscillate with the period of $\sim$ 500 days in phase with the V/R variation.
This implies the quasi-periodic variation of the wing velocities is due to the precession of a one-armed density wave, which also causes the V/R variation.

\section{Discussion}\label{sec:discuss}
\subsection{Summary of the Observational Features}\label{subsec:sum_obs}
As shown in the previous section, the observed line profiles have exhibited remarkable and complicated variabilities in the observing period, not only during the 2009 giant outburst but also after it.
They are summarized as follows:
\begin{enumerate}
	\item\label{fea:profile2009a} The Balmer lines from one orbital cycle before the 2009 giant outburst to one orbital cycle after it are characterized by ``redshifted enhanced component", which had a double or triple peak during the 2009 giant outburst.
	\item\label{fea:profile2009b} Another remarkable feature of the profiles around the giant outburst in 2009 is the blue hump in the Balmer lines.
	\item\label{fea:shoulder} Bright component, or blue ``shoulder", appeared in the blue side of H$\alpha$ line profiles around periastron in 2009 December and 2010 March and October, but not in 2011 September.
	Whenever the blue ``shoulder" showed up, a bright component appeared around similar radial velocity in the H$\beta$ line profiles, although the duration time was not always the same in the H$\alpha$ and the H$\beta$ lines. 
	\item\label{fea:profile2009c} The emission lines exhibited short-term variabilities --- line profile variabilities and the variation of the equivalent width and the peak flux --- in the cycle between the giant outburst in 2009 and the normal outburst in 2010 March.
	\item\label{fea:profile2010} During the normal outburst in 2010 March, the redshifted double-peaked component in the H$\alpha$ and the H$\beta$ line profiles remarkably weakened, while a broad bright component showed up on the blue side of these lines.
	\item\label{fea:profile2011a} No enhanced component nor triple peak appeared in the Balmer lines around the giant outburst in 2011, unlike the giant outburst in 2009.
	However, there were differences in the observed profiles before and after the 2011 giant outburst.
	\item\label{fea:profile2011c} The He I lines exhibited triple-peaked profiles in 2010 October and 2011 September.
	\item\label{fea:EW_E/C} The absolute value of the equivalent width and the peak flux of each line were the highest around the giant outburst in 2009.
In particular those of the Balmer lines showed extremely high values caused by a redshifted enhanced component.
	\item\label{fea:V/Ra} In 2009 August, the V/R ratio of the H$\alpha$ line was quite different ($\ll$ 1) than that expected from 500-day quasi-periodic variation ($>$ 1).
	\item\label{fea:V/Rb} The V/R ratio showed different trends between the Balmer and the He I lines from 2010 August to October --- V/R  $\lesssim$ 1 for the Balmer lines,  while V/R $>$ 1 for the He I lines.
	\item\label{fea:wing} The FWZI of the H$\alpha$ line profile started to decrease by the beginning of 2008, a couple of cycles prior to the start of the decline of the optical brightness.
\end{enumerate}

From here in this section, the structure and evolution of the Be disk indicated by the above properties are discussed.

\subsection{Gas Stream from the Be Disk to the Neutron Star}\label{sec:stream}

\begin{figure*}[t]
 \begin{center}
	\begin{tabular}{c}
	\FigureFile(160mm,200mm){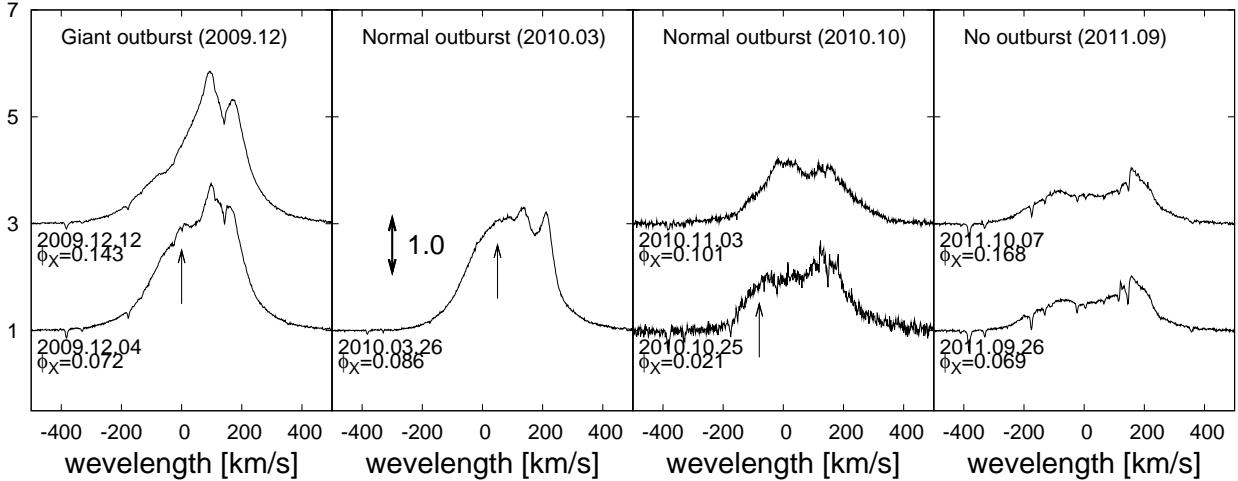} \\
	\end{tabular}
 \end{center}
 \caption{
Comparison of the H$\alpha$ line profiles at two nearby orbital phases after four periastron passages:
From left to right, the profiles are associated with the giant outburst in 2009, the normal outbursts in 2010 March and 2010 October, and no outburst in 2011 September.
The lower profile in each panel is taken at $\phi_X\;<$ 0.1, while the upper one at $\phi_X\;>$ 0.1.
The arrow beneath the lower profile in each of the three left panels denotes the blue ``shoulder".
Note that no profile is available to compare with that on 2010 March 26.
}
 \label{fig:Shoulder}
\end{figure*}

One of the observed variabilities, a bright ``shoulder", appears in the range of $\sim$ $-$100 -- $+$100 $\mathrm{km\;s^{-1}}$ in the H$\alpha$ line profiles after the periastron passage,  $\phi_X\;\sim$  0.1 -- 0.15 [feature (\ref{fea:shoulder})].
Whenever this blue ``shoulder" shows up, the normalized flux in the H$\beta$ line profile becomes higher around a velocity of $\sim$ $-$100 -- 0 $\mathrm{km\;s^{-1}}$ on a different timescale from that of the variation of the blue shoulder.
\citet{Moritani2011} discussed the possibility that the blue shoulder is due to a tidal mass stream from the outermost part of the Be disk.
In this subsection, we examine this possibility in detail using all observed data.

Figure \ref{fig:Shoulder} compares the H$\alpha$ profiles at two nearby orbital phases just after each of several periastron passages; one is taken at $\phi_X\;<$ 0.1 (the lower) and the other at $\phi_X\;\geq$ 0.1 (the upper).
Note that there is no observation just after the periastron passage in the giant outburst in 2011 February, and that we cannot compare line profiles at the periastron in 2012 January because of poor $S/N$.
Note also that the ``shoulder" existed at least until $\phi_X$ = 0.122 in 2010 March (second panel from left), although it can not be determined when the ``shoulder" disappeared because no more observation was carried out in this orbital cycle.
The duration of the appearance of the blue ``shoulder" is the shortest in October 2010 in the three occasions when the ``shoulder" was detected: it seems to have vanished by November 3 ($\phi_X$ = 0.101).
No ``shoulder" was detected after the periastron passage in 2011 September.

The radial velocity of the blue hump [feature (\ref{fea:profile2009b})], $\sim$ $-$100 -- 0 $\mathrm{km\;s^{-1}}$, is almost the same as the velocity of the violet peak of double-peaked profiles showing the V/R variability \citep{Moritani2010}.
This feature indicates that the blue hump arises from the outermost region of the Be disk where the density is higher than usual, probably caused by a one-armed density wave.
The blue ``shoulder" showed up relatively close to the center of the H$\alpha$ line compared to the blue hump (i.e., the radial velocity of the blue ``shoulder" is smaller in the absolute value than that of the blue hump).
This implies that the blue ``shoulder" originates from the region beyond the outer part of the disk emitting the blue hump.
On the other hand, at the orbital phases when the ``shoulder'' shows up, a temporary gas stream is expected to occur, associated with the mass transfer the outer part of the Be disk to the neutron star.
Such a streaming motion has negative line-of-sight velocities, given the observer's direction derived by \citet{Finger1994}.
It is, therefore, likely that the blue ``shoulder" is an observational evidence of the dense gas stream from the Be disk toward the neutron star.
Note that this is consistent with the result that no ``shoulder" was detected after the periastron passage in 2011 September (rightmost panel): no gas stream, hence no mass transfer, therefore no outburst in this periastron passage.
The shortness of the presence of the blue ``shoulder" in 2010 October suggests that the gas flowed for a short time, triggering a short and small outburst (see figure \ref{fig:LC_X}).

\subsection{Warped Be Disk}\label{sec:warp}

\subsubsection{Departure from a Quasi-Keplerian Disk}
Around the giant outburst in 2009, the absolute value of the equivalent width and the peak flux of each line showed the highest values in the last seven years.
As for the H$\alpha$ and the H$\beta$ lines, the redshifted enhanced component was responsible for such a strong emission [feature (\ref{fea:EW_E/C})].
\citet{Grundstrom2006} presented the relationship between the radius of the Be disk and the equivalent width of the H$\alpha$ line based on the numerical model.
\citet{Grundstrom2007} described this relationship in A~$0535+262$ in their figure 4.
On the basis of equation (1) and figure 4 of \citet{Grundstrom2007}, $|EW\mathrm{(H\alpha)}|$ expected for the Be disk with the Roche lobe radius at periastron (5.6 $R_{*}$) ranges between 13 \AA \ ($V$ = 8.9) and 19 \AA \ ($V$= 9.3), where we have used the range of the $V$ magnitude of A~$0535+262$ in the observed period (bottom panel in figure \ref{fig:Wing_Halpha}).
The Be disk is expected to become larger than the Roche lobe when $|EW\mathrm{(H\alpha)}|$ exceed the estimated value.
The estimated values in the case of $V$ $=$ 8.9 and $V$ $=$ 9.3 are denoted by dotted and dashed-dotted lines in the panel (b) of figure \ref{fig:X_EW}, respectively.
Note that some of the stellar parameters adopted by \citet{Grundstrom2007} ($M_{*}\;=20\;M_{\odot}$, $R_{*}\;=\;15R_{\odot}$, $T\;=28000\;\mathrm{K}$) are different from those of this work ($M_{*}\;=25\;M_{\odot}$, $R_{*}\;=\;15R_{\odot}$), but the results are consistent with each other because the relationship mainly depends on the temperature \citep{Grundstrom2006}.

The brightness in $V$-band of A~$0535+262$ remained $\sim$ 8.9 for a couple of years before the double-peaked outburst in 2009 August (JD $\sim$ 2455080).
On the other hand, $|EW\mathrm{(H\alpha)}|$ reached to the level of 13 \AA \ after the double-peaked outburst.
This suggests that the Be disk became wider than the Roche lobe radius at the time.
The Be disk, however, can not grow much beyond the Roche lobe radius because it is truncated by the neutron star \citep{Okazaki2001b,Negueruela2001a} with the truncation radius for $6:1$ resonance as large as the Roche lobe radius in A~$0535+262$.
A~$0535+262$ became fainter in $V$-band after the  double-peaked outburst.
Nevertheless, the high value of $|EW\mathrm{(H\alpha)}|$ indicates that the Be disk radius continued to exceed the Roche lobe radius.

Besides, after the bright ``double-peaked" normal outburst, the V/R ratio deviated far from the 500-day periodic variation (see figure \ref{fig:X_VoverR}): the expected V/R behavior in this period was V $>$ R, while the observed H$\alpha$ and other line profiles exhibited V $\ll$ R [feature (\ref{fea:V/Ra})].
This indicates that a non-axisymmetric bright region has appeared in the disk.

Furthermore, the redshifted enhanced component had a double or triple peak [feature (\ref{fea:profile2009a})].
Some Be stars also have shown triple-peaked profiles: e.g., $\phi$ Per and 59 Cyg.
\citet{Mainz2004} pointed out the possibility that the triple-peaked profiles of these stars are associated with the small absorption by the companion star (subdwarf O-type star).
A~$0535+262$ is, however, the binary of a Be star and a neutron star, and the dip between two adjacent peaks is much broader and deeper than that discussed in \citet{Mainz2004}.
It is therefore unlikely that the neutron star has an influence on the triple-peaked profile by absorption.
In other words, a double or triple peak in the enhanced component originates from the Be disk.

The above three facts (the strong emission, the unexpected V/R behavior and the double or triple peak in the red side) suggest that the Be disk has grown not only in the radial direction but also in the vertical direction in a non-axisymmetric manner --- disk warping.
Consequently, the emission lines reflect the properties of the disk warping probably in addition to the V/R variations via one-armed oscillations.

\subsubsection{Radiatively Driven Warping vs. Tidally Driven Warping}\label{subsec:tidalWarp}
There are two candidate mechanisms for the warping of Be disks: radiatively driven warping (\cite{Pringle1996,Wijers1999}: also see \cite{Porter1998} for an application to Be disks) and tidally driven warping \citep{Lubow2000,Ogilvie2001,Martin2011}.

In \citet{Moritani2011}, we discussed the possibility of the radiatively driven warping \citep{Porter1998}.
We re-estimated the timescale of the precession to be of the order of a week to a month, depending on the stellar parameters.
Although the line profiles and equivalent widths varied on such a short timescale around the 2009 giant outburst (Spectra E -- I, see also figures \ref{fig:X_EW} and \ref{fig:profile_all1}), the global properties of the line profile such as the radial velocity of the enhanced component did not change remarkably.
In fact, the timescale of the observed line profile variability is over several hundred days, which is much longer than the estimated one.
Besides, the triple peak as seen in A~$0535+262$ around the 2009 giant outburst seems to be a unique feature of the Be disk in binary systems, as far as we know from the published H$\alpha$ line profiles \citep{Hanuschik1995,Hanuschik1996,Saad2004,Rivinius2006,Stefl2009,Rivinius2012}.
This, together with the discrepancy between the model and observed variability timescales, might suggest the tidal origin of disk warping.
However, it is premature to rule out the radiative mechanism for warping, given the ambiguity of the radiative model when it is applied to Be disks.

On the other hand, \citet{Martin2011} made numerical models of a viscous decretion Be disk in which the spin of the Be star is misaligned with the orbital axis of the neutron star.
In such a system, the tidal torque by the neutron star has the effect that the Be disk is warped and twisted except in systems with the longest periods.
They also pointed out that it is unlikely that the neutron star will pass through the Be disk in most systems, so that the giant outbursts are probably linked to a warping component.

In any case, the key parameter of the tidally warped Be disk is the tilt angle of the spin axis of the Be star from that of the binary orbit, or angle between the orbital plane and the Be disk.
Indeed, some X-ray binaries are the misaligned systems [\citet{Martin2011, Fragos2010} and reference therein].
The misaligned system is formed when the compact object received the kick velocity at the birth (i.e. asymmetric supernova explosion).
The kick velocity is also expected to make the orbit more eccentric.
\citet{Martin2009} investigated the distribution of the kick velocity in the Be/X-ray binaries, and found that the Maxwellian distributions of the kick velocity is consistent with the observed misalignment or the observed eccentricity, although they could not reproduce both distributions simultaneously.
\citet{Negueruela1999} observed the Be/X-ray binary V~$0332+53$/BQ Cam and found that the optical line profiles have a significant evidence for a tilt between the Be disk plane and the orbital plane.
As for the A~$0535+262$, the maximum of the FWZI of the H$\alpha$ line ($\gtrsim 1000\;\mathrm{km\,s^{-1}}$) suggests the inclination angle of the Be disk is $\sim 30^{\circ}$, slightly different from that of the orbital plane \citep{Finger1994}, if the inner edge of the Be disk reaches to the surface of the photosphere.
It is, however, difficult to constrain the angle between the Be disk plane and the orbital plane, because the optical and the infrared variabilities make it difficult to determine the precise angle of the Be disk (or the rotation axis of the central star) from the line of sight.
Further observations, spectro-polarimetric observations or UV observations in particular, are necessary to check whether A~$0535+262$ is a misaligned system or a coplanar one.

Apart from the warping mechanism itself, another point to be considered on the warping (and precession) of the Be disk is the torque exerted by the central star.
Since the stellar torque tends to keep the disk aligned with the stellar equatorial plane, the Be disk becomes more easily warped when the inner part of the disk is torque-free.
The decrease of the FWZI of the H$\alpha$ line from early 2008 indicates that the inner part of the Be disk was being lost already $\sim$2 yr before the 2009 giant outburst [feature (\ref{fea:wing})].
This fact implies that stellar torque on the Be disk stopped exerting by then, which triggered the precession of the Be disk.
Furthermore, it is likely that without the stellar torque the outer part of the disk is more easily bended towards the orbital plane.
An anti-correlation between the $V$ magnitude and the $|EW\mathrm{(H\alpha)}|$ before the giant outbursts also indicates the disappearance of the inner part of the Be disk, while the outer part still growing \citep{Camero-Arranz2012,Yan2012}.
A similar trend was observed in other systems [e.g., \citet{Stevens1997} for 4U~$1145-619$/V801 Cen; \citet{Yan2012b} for MXB~$0656-072$].
Therefore, it is reasonable to conclude  that the direction of the disk regions bended towards the orbital plane (hereafter, warped regions) is linked to the occurrence of the giant outburst.
The highly redshifted and enhanced asymmetric line profiles [feature (\ref{fea:profile2009a})] indicate that only one of two warped regions can be seen from the observer, while the other warped region extending in the opposite direction with respect to the observer is hidden by the surface of the Be disk.

In the rest of this subsection, we discuss the observed timescale of precession of the warped Be disk and the mass transfer from the warped disk to the neutron star.

\subsubsection{Precession of the Warped Be Disk}\label{subsec:precess}
\begin{figure}[!t]
 \begin{center}
  \FigureFile(80mm,50mm){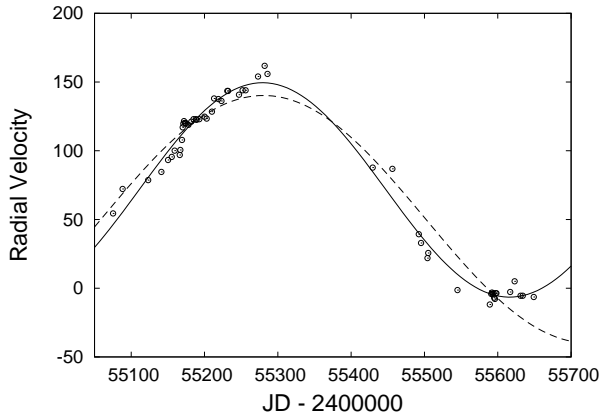}
 \end{center}
 \caption{
Radial velocity of the enhanced component in the H$\alpha$ line profile.
Solid and  dashed lines indicate the result of the fitting for the free-$P_{\mathrm{prec}}$ condition (the best fit value of 674 days) and for a fixed period of 886 days, respectively (see text for detail).
}
 \label{fig:RV enhanced}
\end{figure}

\begin{table*}[!ht]
	\caption{
Fitting results of the radial velocity of the enhancement component using equation (\ref{eq:enhanced}).
}
	\begin{center}
		\begin{tabular}{ccccc} \hline
			$A$ [$\mathrm{km\;s^{-1}}$]	& $\delta$	& $P_{\mathrm{prec}}$ [days]	& $B$ [$\mathrm{km\;s^{-1}}$]	& $\chi^2/d.o.f$ \\ \hline \hline
			$78.0^{-4.0}_{+5.0}$	& $0.09^{-0.02}_{+0.02}$	& $674^{-38}_{+26}$	& $71.5^{-4.0}_{+5.0}$	& 1.4 \\
			$90.0^{-7.5}_{+3.5}$	& $0.01^{-0.01}_{+0.00}$*	& 886 (fixed)	& $50.2^{-2.6}_{+3.6}$	& 0.5 \\ \hline
			\end{tabular}
			
			* : Error cannot be estimated because of rough grid.
	\end{center}
	\label{tbl:Enhanced}
\end{table*}

In order to estimate the precession timescale of the warped Be disk, the radial velocity of the enhanced component in the H$\alpha$ line profile is fitted by a simple sine curve:
\begin{equation}\label{eq:enhanced}
RV_{\mathrm{enhanced}} = A \sin \left[2\pi\left(\frac{t-T_0}{P_{\mathrm{prec}}} +\delta\right)\right] + B,
\end{equation}
where A is the amplitude, $T_0$ is an arbitrary origin of the precession, which we take to be JD 2455050, $\delta$ is the phase offset, $P_{\mathrm{prec}}$ is the period of the precession, and $B$ is the velocity offset from the line center including the systemic velocity.
Note that the systemic velocity has not been determined well so far, and that the velocities from the UV and the optical observations do not agree with each other [\citet{Wang1998} and reference therein].
This is mainly because of the poor orbital coverage for the analysis and the line profile variabilities.

In this analysis the radial velocity of the enhanced component is defined as the central velocity of the Gaussian function that best fits the peak of the enhanced component.
Note that the data around it are not used for the fitting.
Note also that the estimated radial velocity should contain the periodic variability by the orbital motion, but it is ignored, because the amplitude is expected to be less than 10 $\mathrm{km\;s^{-1}}$.
 
 Figure \ref{fig:RV enhanced} shows the observed variation of the radial velocity of the enhanced component and the best fit result, together with the result for a fixed period.
 The result of the best fit (see also table \ref{tbl:Enhanced}) is that $A \; = \; 78.0 \; \mathrm{km\;s^{-1}}$, $P_{\mathrm{prec}} \; = \; 674 \; \mathrm{days}$, $\delta \; = \; 0.09$ and $B \; = \; 71.5 \;\mathrm{km\;s^{-1}}$.
If the $P_{\mathrm{prec}}$ is fixed to the value twice as long as the interval between the giant outbursts in 2009 and 2011, namely $P_{\mathrm{prec}} \; = \; 886 \; \mathrm{days}$, the best fit parameters become $A \; = \; 90.0\;\mathrm{km\;s^{-1}}$,  $\delta \; = \; 0.01$ and $B \; = \; 50.2 \;\mathrm{km\;s^{-1}}$.
These resultant curves are shown in figure \ref{fig:RV enhanced} by the solid ($P_{\mathrm{prec}} \; = \; 674 \; \mathrm{days}$) and the dashed ($P_{\mathrm{prec}} \; = \; 886 \; \mathrm{days}$) lines.

\subsubsection{Mass Transfer from the Precessing Warped Be Disk}
\begin{figure*}[th]
 \begin{center}
\begin{tabular}{cc}
		case(1)	& case(2) \\
		\FigureFile(75mm,50mm){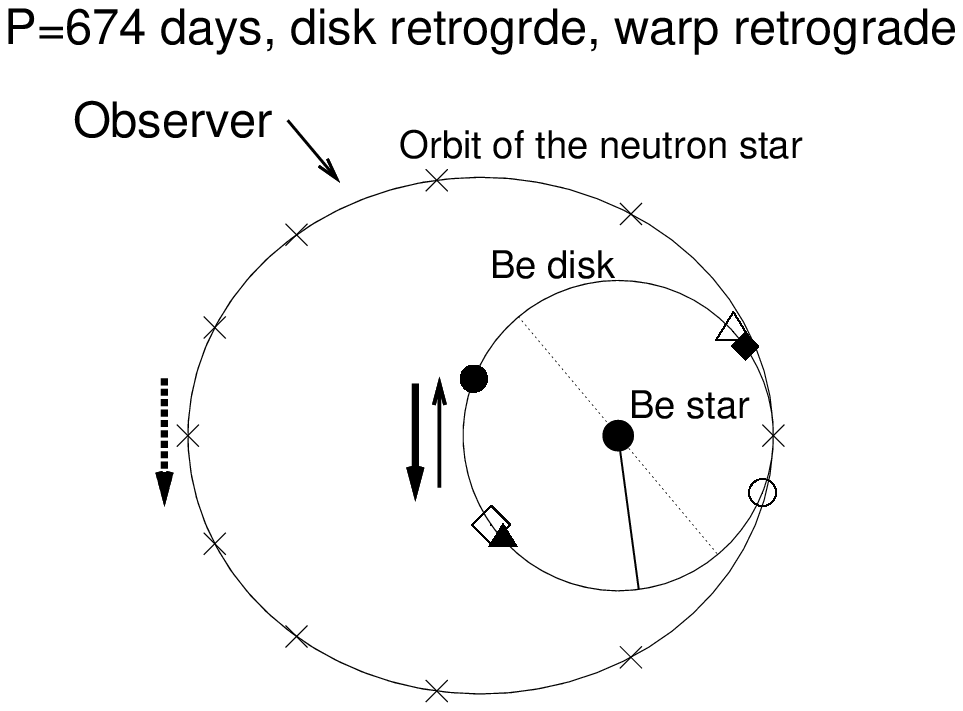} &
		\FigureFile(75mm,50mm){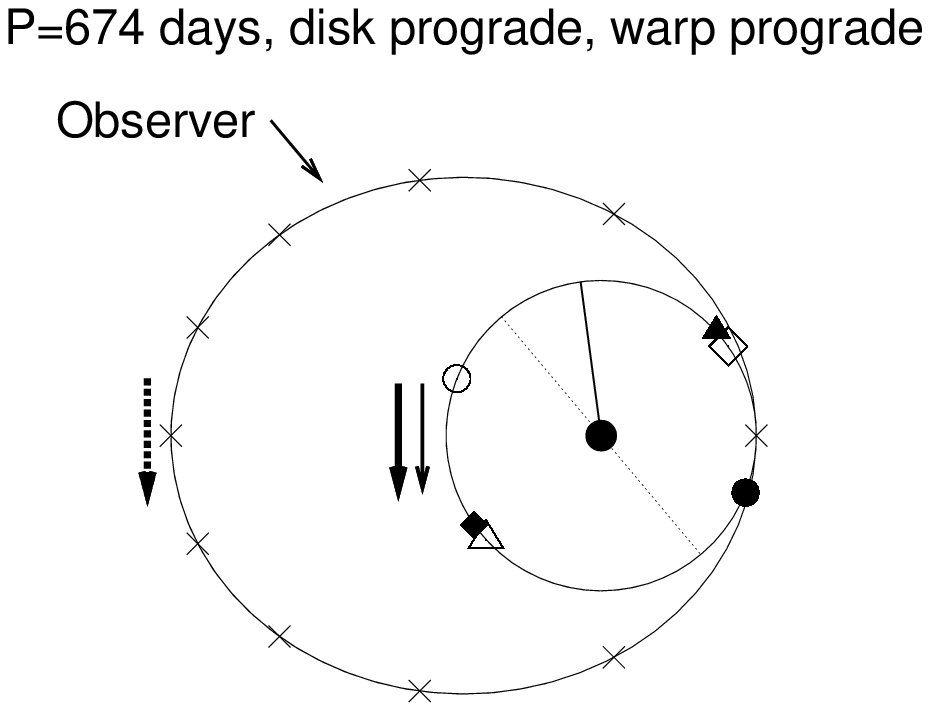} \\
		\FigureFile(75mm,50mm){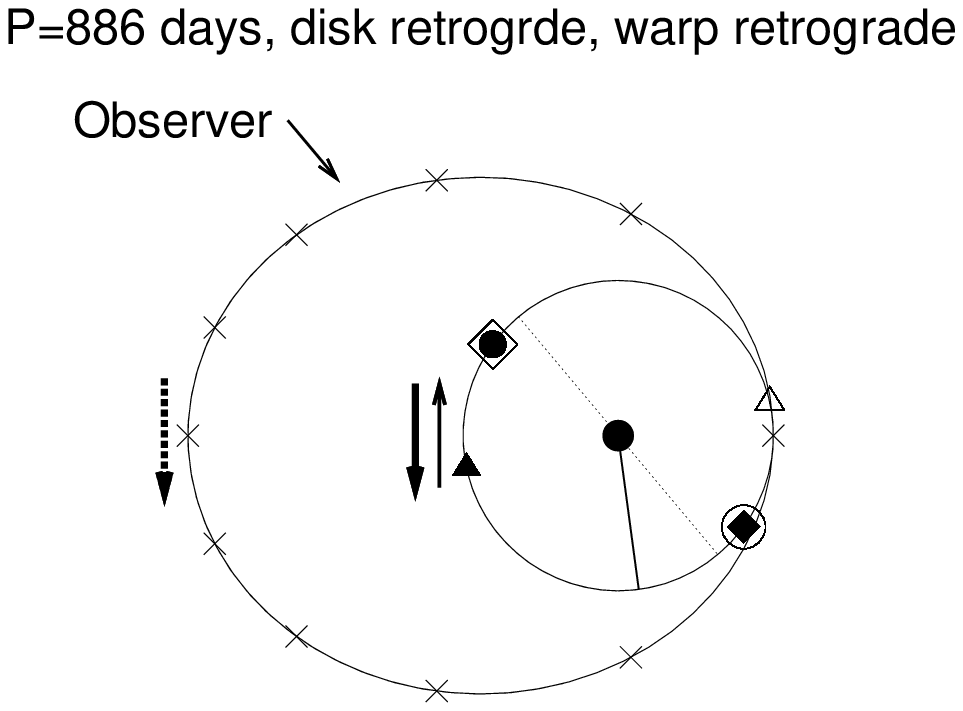} &
		\FigureFile(75mm,50mm){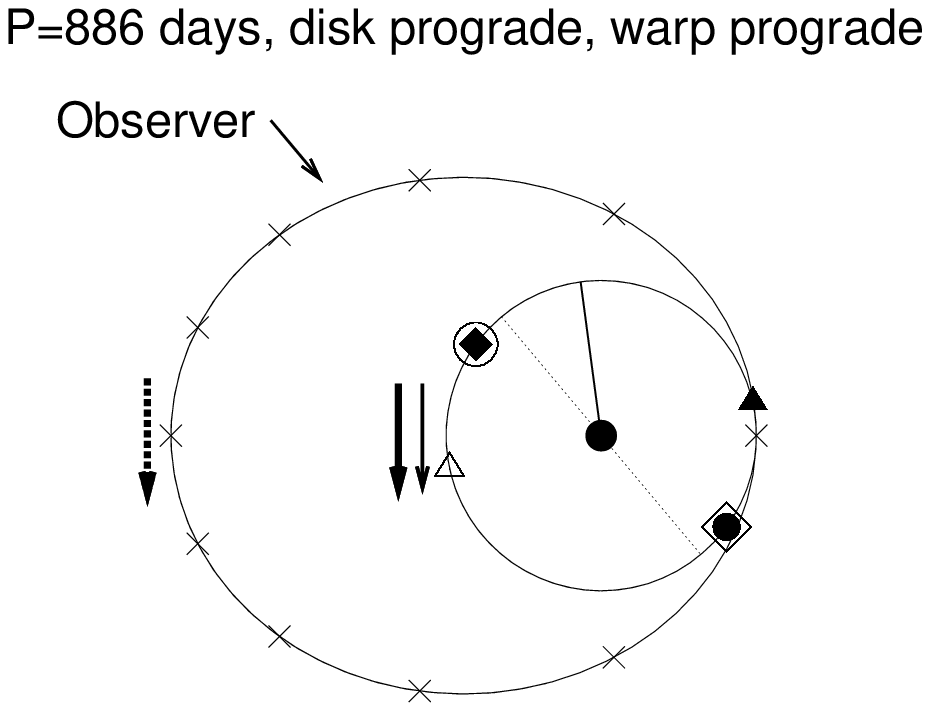} \\
\end{tabular}
\caption{
Expected position of the warped region at the beginning of the giant outbursts in 2009 and 2011 as well as the normal outburst in 2010 March.
The orbital phases with step of 0.1 are marked with crosses on the orbit of the neutron star.
The observer is in the upper-left direction.
Circles, triangles and diamonds indicate the expected positions of two warped regions at the giant outburst in 2009, the normal outbursts in 2010 March and the giant outburst in 2011, respectively.
Open marks denote the positions of the warped region which had emitted the redshifted enhanced component around the 2009 giant outburst, while filled marks show the other warped region on the opposite side.
The dashed line and the solid line denote the isoline of zero radial velocity and the position of the warped region that caused the enhanced component at the origin ($t \; = T_0$), respectively.
The thin and thick solid arrows indicate the rotational and precessional directions of the Be disk, respectively.
Dashed arrows show the direction of the orbit of the neutron star.
}
 \label{fig:Warp_expect}
 \end{center}
\end{figure*}

Using the above results, the expected positions of the enhanced component at several epochs are plotted in figure \ref{fig:Warp_expect}.
The top panels display positions for  $P_{\mathrm{prec}} \; = \; 674 \; \mathrm{days}$ and the bottom panels for $P_{\mathrm{prec}} \; = \; 886 \; \mathrm{days}$.
There are formally four types of geometry with respect to the direction of the precession of warped disk, prograde or retrograde to the rotation of the disk, and the direction of the disk rotation, prograde or retrograde to the orbital motion of the neutron star.
Figure \ref{fig:Warp_expect} shows only plausible cases among them: (1) the warped disk precesses in the retrograde direction and the Be disk also rotates in the retrograde direction (left panels), and (2)  the warped disk precesses in the prograde direction and the Be disk also rotates in the prograde direction (right panels).
Circles, triangles and diamonds indicate the positions on the starting dates of the giant outburst in 2009 (JD 2455164), the normal outbursts in March 2010 (JD 2455283) and the giant outburst in 2011 (JD 2455607), respectively.
The redshifted enhanced component around the 2009 giant outburst originated from the position marked with the open circle.
The other open symbols mark the positions at later times of this particular part of the disk.
Filled marks indicate the shift of the opposite side of the warped Be disk, hidden by the surface of the disk around the 2009 giant outburst.

In case (1), it is likely that during the giant outburst in 2009 the mass transfer occurred from a warped region emitting the redshifted enhanced component [see figure \ref{fig:A0535_history} (c) left].
During the 2011 giant outburst, on the other hand, mass was transferred from the warped region on the other side of the Be disk.
This side was hidden by the surface of the Be disk in front of it, and as a result, no highly enhanced component was observed during the 2011 giant outburst [feature (\ref{fea:profile2011a})].

In case (2), the enhanced mass transfer to the neutron star might have occurred from the hidden warped region during the giant outburst in 2009 [see figure \ref{fig:A0535_history} (c) right].
In this case, during the 2011 giant outburst, the neutron star captured mass from the other warped region which had emitted the redshifted enhanced component one and a half years before.
Feature (\ref{fea:profile2011a}) indicates that this warped region was not so bright in H$\alpha$ as in the 2009 giant outburst, although the blue side of the H$\alpha$ lines temporarily brightened around the 2011 giant outburst.
This is probably because the density of warped region decreased or the precessional angle since the 2009 giant outburst is significantly different from $\pi$ (see the locations of open diamond and circle in the upper right panel of figure \ref{fig:Warp_expect}).

In both cases, the normal outburst in 2010 March was very bright probably because one of the warped regions still lied near the periastron.
Note that the ``double-peaked" normal outburst in 2009 August can be explained similarly.
At the time, the warped region seems to have already around the periastron.
Therefore, the former peak before the periastron (JD 2455048, $\phi_X$ = 0.96) was probably connected to the warped region, while the latter peak around periastron was caused by the mass transfer from a different, unwarped part of the disk, as in the case of other normal outbursts.

\subsection{Structural Change of the Be Disk and its Relationship to the X-ray Activity in Recent Years}
\begin{figure*}[p]
 \begin{center}
  \FigureFile(165mm,80mm){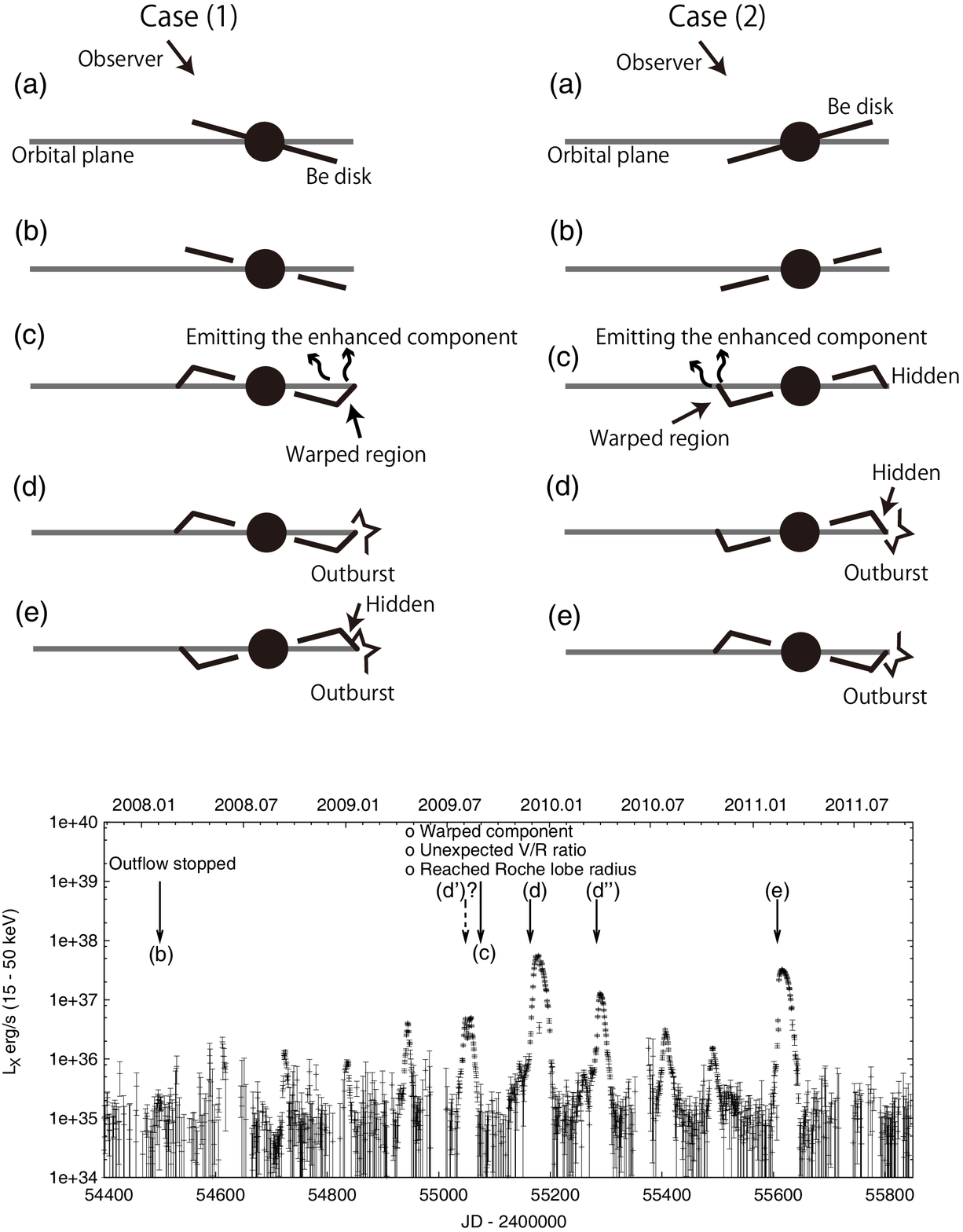}
 \end{center}
 \caption{
Evolution of the Be disk in A~$0535+262$ for the last seven years.
Schematic pictures on the left side are for case (1), while those on the right side are for case (2).
X-ray light curve from 2008 to 2011 is also plotted with arrows indicating events at corresponding time. 
}
 \label{fig:A0535_history}
\end{figure*}

For the last seven years, A~$0535+262$ exhibited various photometric and spectroscopic variabilities caused by the activity of the Be disk, as well as the high activity in X-rays.
The evolution of the Be disk in recent years clarified by this work can be summarized as follows (figure \ref{fig:A0535_history}):

\begin{enumerate}
	\item After the giant outburst in May 2005 and subsequent two normal outbursts, A~$0535+262$ went into a two-year quiescent state.
	The Be disk did not disappear, in contrast to the 1994 giant outburst [state (a) of figure  \ref{fig:A0535_history}].
	\item In the quiescent state, the Be star had a quasi-Keplerian disk exhibiting the periodic V/R variation with a period of 500 days via the one-armed oscillation.
	\item By the beginning of 2008, the mass ejection from the central star stopped and hence the torque exerted by the central star on the innermost part of the Be disk disappeared [state (b)].
	The density of the inner region of the disk decreased, resulting in the narrower wing of the H$\alpha$ line.
	\item Without the competing  stellar torque, the Be disk is warped and precesses more easily by the radiative torque from the central star or the tidal torque from the neutron star [state (c)].
	\item By August 2009, the warping of the Be disk grew to exhibit an enhanced emission.
	One of two warped regions happened to lie near the periastron causing the ``double-peaked" normal outburst in 2009 July/August [state (d')].
	\item One cycle later, the same warped region precessed and reached very close to the neutron star orbit around the periastron [state (d)].
	A large amount of mass was transferred from the warped Be disk to the neutron star, which caused the giant outburst in 2009.
	\item Around the normal outburst in 2010 March, the warped region still lay near the periastron, supplying a larger amount of gas than normal, so that the X-ray flux was high [state (d'')].
	The short-term variability detected from 2010 January to March might be caused by the interaction between the warped Be disk and the neutron star.
	\item After that, the normal outbursts occurred at the periastron passages via mass transfer from the outer region of the Be disk.
	\item When the other warped region on the opposite side came near the periastron in 2011 February, the giant outburst occurred again [state (e)].
	No enhanced component was seen probably because the warped region was not no bright in H$\alpha$ or mostly hidden by the surface of the Be disk.
	\item  Since the enormous amount of gas was removed from the Be disk in the X-ray active phase, the disk became tenuous and small, emitting a weak emission.
After the giant outburst in 2011, the Be disk returned to a quasi-Keplerian disk, which was smaller than the Roche lobe radius of the Be star at periastron.
	With no sufficient mass transfer from the disk, A~$0535+262$ settled into the quiescent state in X-rays. 
\end{enumerate}

It is still disputable when the Be disk starts warping, because \citet{Martin2011} suggested that the Be disk is warped even if the inner edge in contact with to the surface of the star, although the effect of warping is expected to be small in this case.
The effect of the mass ejection from the photosphere as well as the detailed relationship between the growth and dissipation of the Be disk and X-ray activity will be discussed in Okazaki et al. (2013, in prep.).

\subsection{Open Issues}\label{sec:open}
Many observed features indicate that the giant outburst is linked to the mass transfer from the precessing, warped Be disk to the neutron star.
However, there are many open issues in the Be disk evolution as well as the interaction between the Be disk and the neutron star.
First of all, there are currently no crucial clues to the warping mechanism nor to the direction of precession of the warped Be disk, prograde or retrograde.
If the direction of precession is determined, the origin of the warping becomes clear and vice versa, because the precessing direction depends on the mechanism.
If the Be disk is warped by the radiative torque, the precession is prograde.
On the other hand, the precession of the tidally warped Be disk is retrograde.
\citet{Caproni2006} investigated the direction of precession of accretion disks in X-ray binaries and Active Galactic Nuclei and found that for X-ray binaries both mechanisms predict consistent precession periods with the observed data.
In A~$0535+262$, the warping mechanism can be constrained by estimating precession period for both mechanisms using realistic conditions.
The direction of precession can also be determined by comparing the profile variabilities in various lines.
We will preform these detailed study in subsequent papers.
If interferometric observations becomes available for this star ($V$ $\sim$ 9), they will also be able to resolve the structure of the Be disk directly and hence the direction of the precession, when a giant outburst occurs in the future [e.g., \citet{Pott2010} for 48 Lib].

Next, the angle between the Be disk plane and the orbital plane is still unknown (see section \ref{sec:warp}).
Spectro-polarimetric or UV observations are effective to constrain the inclination of the Be star, or to check whether A~$0535+262$ is a misaligned or coplanar system.
The information about the inclination of the Be star can also give a hint concerning the precessing direction of the warped Be disk because the sequence of the X-ray activity is different between the prograde and the retrograde cases (figure  \ref{fig:A0535_history}).

Moreover, the short-term variabilities observed from January to March 2010 [feature (\ref{fea:profile2009c})] suggest a complicated interaction between the warped Be disk and the neutron star, of which the details are far from clear yet.
Also the triple-peaked profiles in the He I lines [feature (\ref{fea:profile2011c})] are probably linked to the complicated structure of the Be disk.
Further study is needed to deal with these variabilities and profiles.

Finally, the structure of the Be disk, particularly the evolution of the density distribution, is not well constrained by the spectroscopic observations only.
Spectro-polarimetric observations can be used to probe the radial density distribution of the Be disk \citep{Draper2011}.
\citet{Draper2011} monitored two Be stars losing mass from the disk and found that the polarization across the Balmer jump can be a tracer of the innermost disk density, whereas the $V$-band polarization can trace the total disk mass.
Therefore, the polarimetric observation can reveal when the mass ejection from the star stops and the Be disk starts precession/warping.

After the giant outburst in 1994, the Be disk disappeared and reformed again two years later \citep{Clark1998a,Grundstrom2007}.
As for the 2009 and 2011 giant outbursts, in spite of a current quiescent state of A~$0535+262$, the existing emission lines indicate the presence of the Be disk.
The small equivalent width of these lines and the small line width of the H$\alpha$ line, however, suggest that the Be disk is the weakest in the last seven years with the tenuous inner region.
Recent gradual increase of FWZI of the H$\alpha$ line suggests that the mass ejection from the star started again.
Therefore, frequent observations of the Be disk in A~$0535+262$ for the next few years can determine whether the Be disk will disappear or grow again, and give more clues to the relationship between the Be disk evolution and the X-ray activity.

\section{Conclusion}\label{sec:conclude}
The Be/X-ray binary A~$0535+262$ was X-ray active from 2008 to 2011, exhibiting several normal outbursts and two giant outbursts during this period.
Optical high-dispersion spectroscopic observations of A~$0535+262$ were carried out from 2009 April to 2012 April, including the 2009 and 2011 giant outbursts, in order to investigate the Be disk structure in the X-ray active state.
In particular, our observations covered the whole cycle, from the beginning to the end, of the giant outburst in 2009 for the first time with high-dispersion spectrographs.

The observed emission lines exhibited various variabilities, reflecting the change of the structure of the Be disk.
Among the observed variable features, particularly important as probes for the interaction with the neutron star are the blue ``shoulder'' and the redshifted enhanced component that characterize the H$\alpha$ line profiles in this X-ray active period.

The bright blue ``shoulder" was detected after the periastron passage associated with several outbursts including the 2009 giant outburst.
This indicates that this feature originates from the dense gas stream from the outermost part of the Be disk to the neutron star at periastron, triggering the outburst.

On the other hand, the strong emission line profiles during the giant outburst in 2009 imply conspicuous active components in the Be disk.
The redshifted enhanced component, which occasionally had a triple peak and caused an unexpected behavior of the V/R ratio, gives an evidence for a warped Be disk.
We analysed the FWZI of the H$\alpha$ line profiles and found that the Be star stopped ejecting mass by the beginning of 2008.
After that, the Be disk seems to have become warped and started precession.
The warping of the Be disk grew to emit the redshifted enhanced component in the emission lines by the time of the double-peaked normal outburst in August 2009.

We estimated the timescale of the precession of the warped Be disk to be several hundred days.
Using the derived precession parameters, we found that the region in the Be disk bended towards the orbital plane was near the periastron at the beginning of the 2009 and 2011 giant outbursts as well as the normal outburst in 2010 March.
It is, therefore, likely that the giant outbursts occurred in consequence of the enhanced mass transfer from the precessing warped Be disk to the neutron star, although the direction of the precession remains disputable.

Our seven-year monitoring of A~$0535+262$ with high-dispersion spectrographs revealed the evolution of the Be disk and its relationship with the X-ray activity.
The results strongly suggest that the giant outburst occur when the neutron star captures a large amount of gas from the warped Be disk.
Although the current study provided an insight on the long-term structural change of the Be disk and its relationship to the X-ray activity, there are still many open questions.
Further study, both theoretical and observational, is necessary to establish the relationship between the Be disk evolution and the X-ray activity, solving these questions.
It is also important to study whether such a relationship applies to other Be/X-ray binaries as well.
In particular, multi-wavelength observations with various methods (photometry, spectroscopy, polarimetry etc.) will be useful for these purposes.

Since the giant outburst in 2011, the Be disk in A~$0535+262$ has been in a weak state.
According to the X-ray activity since its discovery \citep{Camero-Arranz2012}, A~$0535+262$ seems to return to the X-ray active phase in less than ten years unless the Be disk dissipates.
It is therefore important to monitor A~$0535+262$ in preparation to the next X-ray active phase.


\bigskip
We thank the anonymous referee for constructive comments.
We are very grateful to Dr. Bun'ei Sato for kindly observing A~$0535+262$.
This work was supported by Research Fellowships for the Promotion of Science for Young Scientists (YM).
We also acknowledge support of JSPS Grant-in-Aid for Scientific Research (20540240, 18340055, 24540235), the Grant-in-Aid for the Global COE Program "The Next Generation of Physics, Spun from Universality and Emergence" from the Ministry of Education, Culture, Sports, Science and Technology (MEXT) of Japan, and the Collaborative Research Program 2010, Information Initiative Center, Hokkaido University.


\appendix
\section{All Obtained Line Profiles and Quantities Characterizing Profiles}\label{app:obs}
\begin{figure*}[]
 \begin{center}
  \FigureFile(160mm,250mm){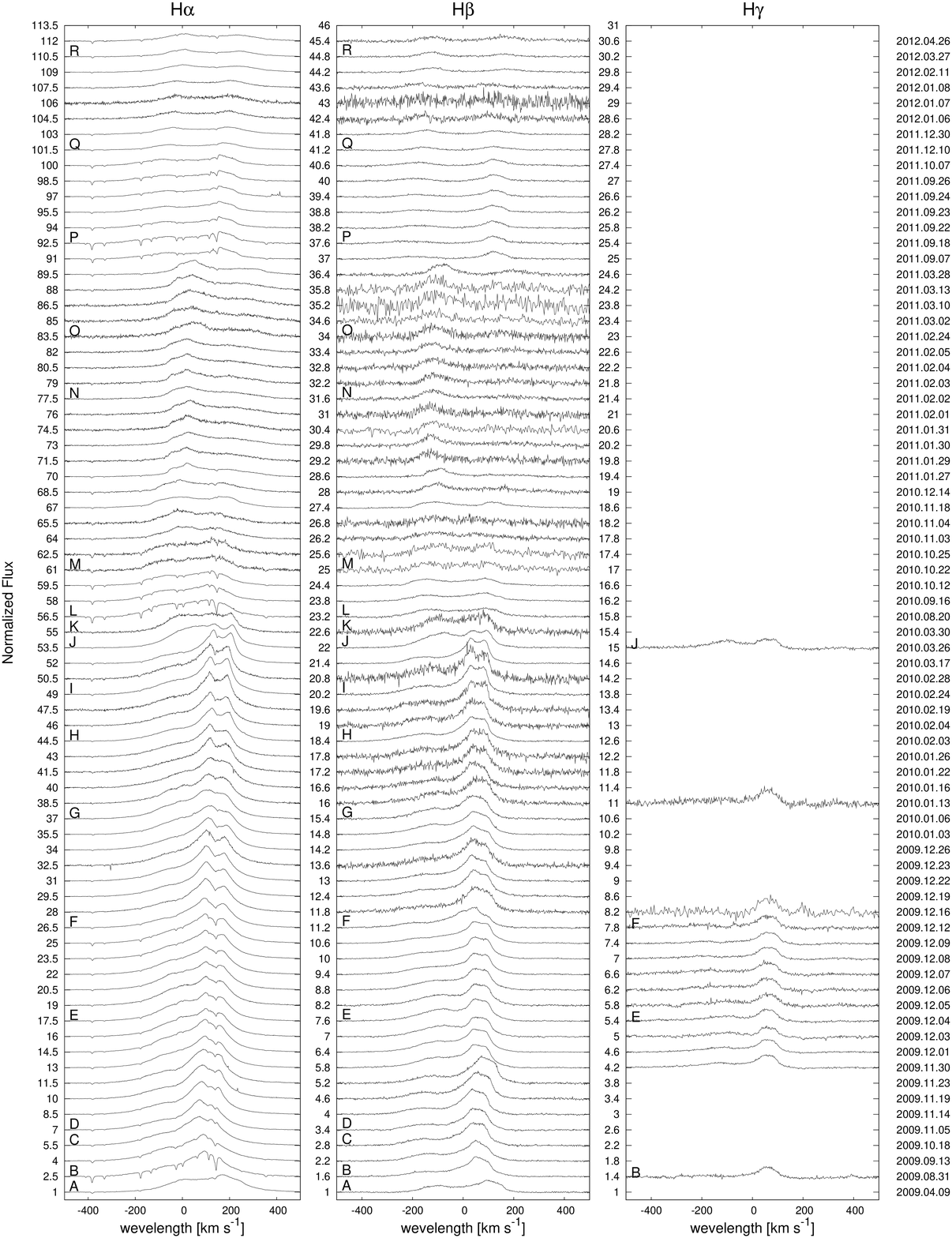}
 \end{center}
 \caption{
All observed Balmer line profiles: the H$\alpha$, the H$\beta$ and the H$\gamma$ line profiles from left to right.
The observed date is annotated on the right side of the figure.
Alphabets above profiles denote representative spectra in figures \ref{fig:profile_rep} -- \ref{fig:profile_rep3}.
}
 \label{fig:profile_all1}
\end{figure*}

\begin{figure*}[]
 \begin{center}
  \FigureFile(160mm,250mm){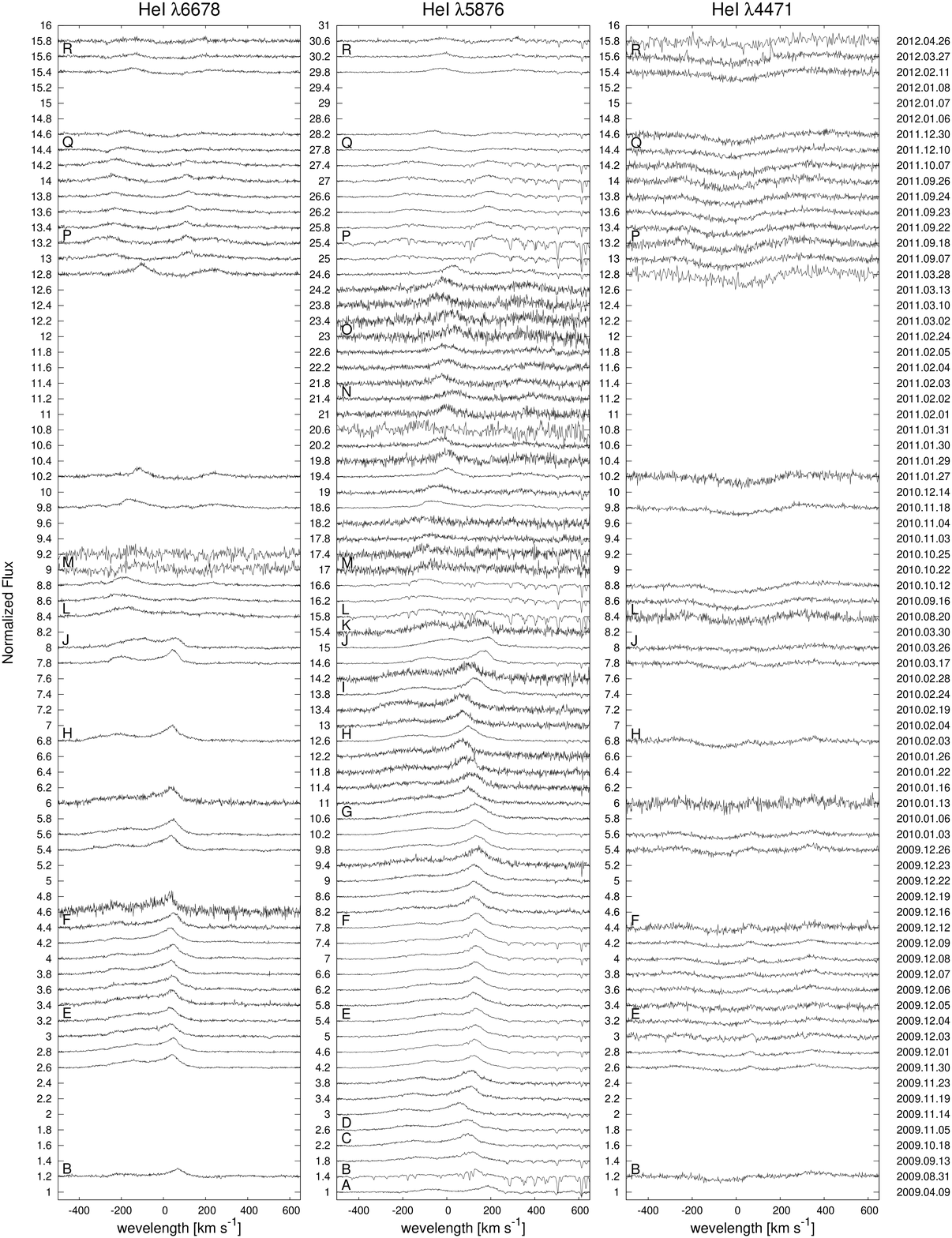}
 \end{center}
 \caption{
Same as figure \ref{fig:profile_all1}, but for all observed He I line profiles.
From left to right, the He I $\lambda$6678, the He I $\lambda$5876 and the He I $\lambda$4471 line profiles.
}
 \label{fig:profile_all2}
\end{figure*}

Below we show all obtained Balmer and He I line profiles in figures \ref{fig:profile_all1} and \ref{fig:profile_all2}, respectively.
The observational date are written on the right side of the each figure.
The alphabets above the profiles denote the representative profiles displayed in figures \ref{fig:profile_rep}, \ref{fig:profile_rep2} and \ref{fig:profile_rep3}.
Equivalent widths and V/R ratios of each line are summarized in table \ref{tbl:Obs_results}, and wing velocities of H$\alpha$ line profile are listed in table \ref{tbl:Obs_wing}.

\begin{longtable}[]{ccl*{6}{r}}
	\caption{
Equivalent width in unit of \AA \ and the V/R ratio of each line profile.
Columns 1 and 2 list the date and HJD, respectively.
The outburst phase $\phi_X$ by \citet{Moritani2010} is also shown in column 2.
The word ``n/a" as the V/R values means that the value cannot be estimated because the profile has more than two peaks, or the normalized peak is weaker than continuum ($<$ 1).
The mark ``$\times$" indicate that the value can not be determined because of the poor $S/N$.
}
	\label{tbl:Obs_results}\\
	\hline
	 Date	& HJD ($\phi_X$)	&	& H$\alpha$	& H$\beta$	& H$\gamma$	& He I $\lambda$6678	& He I $\lambda$5876	& He I $\lambda$4471	\\
	\hline \hline
	\endhead
	\endfoot
	\endlastfoot
2009.04.09	& 2454930.955	& EW [\AA]	& $-$11.35	& $-$1.53	& $-$	& $-$	& $-$0.48	& $-$	\\
	& (0.902)	& V/R	& 0.75	& 0.57	& $-$	& $-$	& 0.49	& $-$	\\ \hline
2009.08.31	& 2455075.264	& EW [\AA]	& $-$12.84	& $-$2.13	& $-$0.19	& $-$0.38	& $-$0.15	& 0.02	\\
	& (0.211)	& V/R	& n/a	& 0.26	& 0.15	& 0.37	& 0.30	& n/a	\\ \hline
2009.09.13	& 2455088.242	& EW [\AA]	& $-$14.91	& $-$2.30	& $-$	& $-$	& $-$0.92	& $-$	\\
	& (0.329)	& V/R	& n/a	& 0.37	& $-$	& $-$	& 0.48	& $-$	\\ \hline
2009.10.18	& 2455123.143	& EW [\AA]	& $-$16.60	& $-$2.49	& $-$	& $-$	& $-$1.19	& $-$	\\
	& (0.645)	& V/R	& n/a	& 0.32	& $-$	& $-$	& 0.44	& $-$	\\ \hline
2009.11.05	& 2455141.131	& EW [\AA]	& $-$14.80	& $-$2.20	& $-$	& $-$	& $-$1.17	& $-$	\\
	& (0.808)	& V/R	& n/a	& 0.27	& $-$	& $-$	& 0.46	& $-$	\\ \hline
2009.11.14	& 2455150.289	& EW [\AA]	& $-$16.68	& $-$2.63	& $-$	& $-$	& $-$1.25	& $-$	\\
	& (0.891)	& V/R	& n/a	& 0.26	& $-$	& $-$	& 0.47	& $-$	\\ \hline
2009.11.19	& 2455155.255	& EW [\AA]	& $-$17.63	& $-$2.88	& $-$	& $-$	& $-$1.31	& $-$	\\
	& (0.936)	& V/R	& n/a	& 0.26	& $-$	& $-$	& 0.45	& $-$	\\ \hline
2009.11.23	& 2455159.301	& EW [\AA]	& $-$17.83	& $-$2.92	& $-$	& $-$	& $-$1.31	& $-$	\\
	& (0.973)	& V/R	& n/a	& 0.30	& $-$	& $-$	& 0.51	& $-$	\\ \hline
2009.11.30	& 2455166.190	& EW [\AA]	& $-$17.03	& $-$3.00	& $-$0.86	& $-$0.80	& $-$1.46	& 0.01	\\
	& (0.036)	& V/R	& n/a	& 0.47	& 0.40	& 0.59	& 0.58	& n/a	\\ \hline
2009.12.01	& 2455167.084	& EW [\AA]	& $-$18.36	& $-$3.06	& $-$0.81	& $-$0.99	& $-$1.47	& 0.23	\\
	& (0.044)	& V/R	& n/a	& 0.50	& 0.39	& 0.59	& 0.56	& n/a	\\ \hline
2009.12.03	& 2455169.277	& EW [\AA]	& $-$16.24	& $-$2.55	& $-$0.45	& $-$0.89	& $-$1.41	& 0.03	\\
	& (0.064)	& V/R	& n/a	& 0.59	& 0.48	& 0.65	& 0.62	& n/a	\\ \hline
2009.12.04	& 2455170.229	& EW [\AA]	& $-$16.39	& $-$2.95	& $-$0.74	& $-$0.88	& $-$1.41	& 0.09	\\
	& (0.072)	& V/R	& n/a	& 0.65	& 0.44	& 0.59	& 0.62	& n/a	\\ \hline
2009.12.05	& 2455171.088	& EW [\AA]	& $-$17.23	& $-$2.88	& $-$0.65	& $-$1.26	& $-$1.48	& 0.33	\\
	& (0.080)	& V/R	& n/a	& 0.52	& 0.40	& n/a	& 0.53	& n/a	\\ \hline
2009.12.06	& 2455172.033	& EW [\AA]	& $-$16.85	& $-$2.84	& $-$0.68	& $-$0.97	& $-$1.47	& $-$0.02	\\
	& (0.089)	& V/R	& n/a	& 0.46	& n/a	& 0.49	& 0.48	& n/a	\\ \hline
2009.12.07	& 2455173.311	& EW [\AA]	& $-$16.11	& $-$2.69	& $-$0.86	& $-$0.94	& $-$1.36	& $-$0.01	\\
	& (0.100)	& V/R	& n/a	& n/a	& 0.41	& 0.45	& 0.41	& n/a	\\ \hline
2009.12.08	& 2455174.150	& EW [\AA]	& $-$16.32	& $-$2.73	& $-$0.69	& $-$0.87	& $-$1.32	& 0.31	\\
	& (0.108)	& V/R	& n/a	& n/a	& 0.27	& 0.38	& 0.37	& n/a	\\ \hline
	&	&	&	&	&	&	&	& \\
2009.12.09	& 2455175.060	& EW [\AA]	& $-$16.30	& $-$2.70	& $-$0.43	& $-$1.04	& $-$1.32	& 0.02	\\
	& (0.116)	& V/R	& n/a	& n/a	& 0.18	& 0.51	& 0.31	& n/a	\\ \hline
2009.12.12	& 2455178.107	& EW [\AA]	& $-$15.77	& $-$2.52	& $-$0.69	& $-$0.83	& $-$1.48	& $-$0.04	\\
	& (0.143)	& V/R	& n/a	& n/a	& 0.24	& 0.43	& 0.34	& n/a	\\ \hline
2009.12.16	& 2455181.951	& EW [\AA]	& $-$16.96	& $-$3.02	& $-$0.41	& $-$1.10	& $-$1.65	& $\times$	\\
	& (0.179)	& V/R	& n/a	& n/a	& n/a	& 0.52	& 0.36	& $\times$	\\ \hline
2009.12.19	& 2455185.060	& EW [\AA]	& $-$17.60	& $-$3.24	& $-$	& $-$	& $-$1.73	& $-$	\\
	& (0.207)	& V/R	& n/a	& n/a	& $-$	& $-$	& 0.37	& $-$	\\ \hline
2009.12.22	& 2455188.223	& EW [\AA]	& $-$17.98	& $-$2.90	& $-$	& $-$	& $-$1.93	& $-$	\\
	& (0.235)	& V/R	& n/a	& 0.29	& $-$	& $-$	& 0.37	& $-$	\\ \hline
2009.12.23	& 2455189.126	& EW [\AA]	& $-$18.15	& $-$2.82	& $-$	& $-$	& $-$2.10	& $-$	\\
	& (0.244)	& V/R	& n/a	& 0.31	& $-$	& $-$	& 0.43	& $-$	\\ \hline
2009.12.26	& 2455192.903	& EW [\AA]	& $-$16.82	& $-$2.94	& $-$	& $-$0.74	& $-$1.61	& $-$0.07	\\
	& (0.278)	& V/R	& n/a	& 0.32	& $-$	& 0.38	& 0.39	& n/a	\\ \hline
2010.01.03	& 2455200.104	& EW [\AA]	& $-$16.68	& $-$3.01	& $-$	& $-$0.80	& $-$1.67	& 0.08	\\
	& (0.343)	& V/R	& n/a	& 0.46	& $-$	& 0.42	& 0.46	& n/a	\\ \hline
2010.01.06	& 2455202.960	& EW [\AA]	& $-$17.36	& $-$2.97	& $-$	& $-$	& $-$1.83	& $-$	\\
	& (0.369)	& V/R	& n/a	& 0.49	& $-$	& $-$	& 0.49	& $-$	\\ \hline
2010.01.13	& 2455210.120	& EW [\AA]	& $-$16.38	& $-$3.15	& $-$0.69	& $-$1.04	& $-$1.709	& 0.05	\\
	& (0.434)	& V/R	& n/a	& 0.48	& 0.35	& 0.46	& 0.46	& n/a	\\ \hline
2010.01.16	& 2455213.053	& EW [\AA]	& $-$16.84	& $-$2.71	& $-$	& $-$	& $-$1.65	& $-$	\\
	& (0.461)	& V/R	& n/a	& 0.41	& $-$	& $-$	& 0.43	& $-$	\\ \hline
2010.01.22	& 2455219.040	& EW [\AA]	& $-$17.03	& $-$2.82	& $-$	& $-$	& $-$1.53	& $-$	\\
	& (0.515)	& V/R	& n/a	& n/a	& $-$	& $-$	& 0.46	& $-$	\\ \hline
2010.01.26	& 2455223.151	& EW [\AA]	& $-$16.04	& $-$3.13	& $-$	& $-$	& $-$2.07	& $-$	\\
	& (0.552)	& V/R	& n/a	& n/a	& $-$	& $-$	& 0.46	& $-$	\\ \hline
2010.02.03	& 2455231.072	& EW [\AA]	& $-$16.55	& $-$2.80	& $-$	& $-$1.01	& $-$1.64	& 0.14	\\
	& (0.624)	& V/R	& n/a	& n/a	& $-$	& 0.49	& 0.41	& n/a	\\ \hline
2010.02.04	& 2455232.133	& EW [\AA]	& $-$16.11	& $-$2.62	& $-$	& $-$	& $-$1.36	& $-$	\\
	& (0.634)	& V/R	& n/a	& n/a	& $-$	& $-$	& 0.43	& $-$	\\ \hline
2010.02.19	& 2455247.045	& EW [\AA]	& $-$18.23	& $-$3.14	& $-$	& $-$	& $-$1.95	& $-$	\\
	& (0.769)	& V/R	& n/a	& n/a	& $-$	& $-$	& 0.57	& $-$	\\ \hline
2010.02.24	& 2455251.965	& EW [\AA]	& $-$18.86	& $-$3.23	& $-$	& $-$	& $-$1.93	& $-$	\\
	& (0.814)	& V/R	& n/a	& n/a	& $-$	& $-$	& 0.44	& $-$	\\ \hline
	&	&	&	&	&	&	&	& \\
2010.02.28	& 2455256.035	& EW [\AA]	& $-$17.09	& $-$3.31	& $-$	& $-$	& $-$1.90	& $-$	\\
	& (0.851)	& V/R	& n/a	& n/a	& $-$	& $-$	& 0.44	& $-$	\\ \hline
2010.03.17	& 2455272.960	& EW [\AA]	& $-$14.78	& $-$2.59	& $-$	& $-$0.66	& $-$1.23	& 0.11	\\
	& (0.004)	& V/R	& n/a	& n/a	& $-$	& 0.51	& 0.50	& n/a	\\ \hline
2010.03.26	& 2455281.982	& EW [\AA]	& $-$15.26	& $-$2.91	& $-$0.63	& $-$0.82	& $-$1.55	&$-$0.05 	\\
	& (0.086)	& V/R	& n/a	& n/a	& 0.81	& 1.01	& 0.88	& n/a	\\ \hline
2010.03.30	& 2455285.925	& EW [\AA]	& $-$12.61	& $-$2.79	& $-$	& $-$	& $-$1.51	& $-$	\\
	& (0.122)	& V/R	& n/a	& n/a	& $-$	& $-$	& 0.86	& $-$	\\ \hline
2010.08.20	& 2455429.309	& EW [\AA]	& $-$10.44	& $-$1.58	& $-$	& $-$0.91	& $-$0.83	& 0.18	\\
	& (0.422)	& V/R	& 0.78	& 0.87	& $-$	& 2.34	& 2.64	& n/a	\\ \hline
2010.09.16	& 2455456.298	& EW [\AA]	& $-$10.16	& $-$1.44	& $-$	& $-$0.62	& $-$0.80	& 0.34	\\
	& (0.667)	& V/R	& 0.68	& 0.75	& $-$	& 2.03	& 1.67	& n/a	\\ \hline
2010.10.12	& 2455482.212	& EW [\AA]	& $-$8.97	& $-$1.26	& $-$	& $-$0.57	& $-$0.74	& 0.11	\\
	& (0.902)	& V/R	& 0.69	&0.89 	& $-$	& 2.37	& 2.35	& n/a	\\ \hline
2010.10.22	& 2455492.219	& EW [\AA]	& $-$8.50	& $-$1.33	& $\times$	& $-$0.29	& $-$0.58	& $\times$	\\
	& (0.993)	& V/R	& 0.77	& 0.99	& $\times$	& n/a	& n/a	& $\times$	\\ \hline
2010.10.25	& 2455495.328	& EW [\AA]	& $-$8.86	& $-$1.60	& $\times$	& $-$0.05	& $-$0.73	& $\times$	\\
	& (0.021)	& V/R	& 0.69	& 0.98	& $\times$	& n/a	& n/a	& $\times$	\\ \hline
2010.11.03	& 2455504.075	& EW [\AA]	& $-$8.09	& $-$1.14	& $-$	& $-$	& $-$0.16	& $-$	\\
	& (0.101)	& V/R	& 1.07	& 1.15	& $-$	& $-$	& n/a	& $-$	\\ \hline
2010.11.04	& 2455505.187	& EW [\AA]	& $-$7.85	& $-$0.74	& $-$	& $-$	& $-$0.51	& $-$	\\
	& (0.111)	& V/R	& 1.17	& 0.96	& $-$	& $-$	& n/a	& $-$	\\ \hline
2010.11.18	& 2455519.187	& EW [\AA]	& $-$8.87	& $-$1.03	& $-$	& $-$0.59	& $-$0.84	& 0.20	\\
	& (0.238)	& V/R	& 0.97	& 1.04	& $-$	& 2.30	& 2.35	& n/a	\\ \hline
2010.12.14	& 2455544.966	& EW [\AA]	& $-$7.76	& $-$1.27	& $-$	& $-$	& $-$0.36	& $-$	\\
	& (0.472)	& V/R	& 1.33	& 1.68	& $-$	& $-$	& 2.27	& $-$	\\ \hline
2011.01.27	& 2455589.165	& EW [\AA]	& $-$7.13	& $-$0.73	& $-$	& $-$0.28	& $-$0.68	& 0.18	\\
	& (0.872)	& V/R	& 1.83	& 3.82	& $-$	& 2.26	& 2.33	& n/a	\\ \hline
2011.01.29	& 2455591.075	& EW [\AA]	& $-$7.43	& $-$0.85	& $-$	& $-$	& $-$0.67	& $-$	\\
	& (0.890)	& V/R	& 1.69	& 2.83	& $-$	& $-$	& 2.20	& $-$	\\ \hline
2011.01.30	& 2455592.110	& EW [\AA]	& $-$7.70	& $-$1.03	& $-$	& $-$	& $-$0.41	& $-$	\\
	& (0.899)	& V/R	& 1.76	& 3.35	& $-$	& $-$	& 2.58	& $-$	\\ \hline
2011.01.31	& 2455592.992	& EW [\AA]	& $-$7.31	& $-$0.25	& $-$	& $-$	& $-$0.08	& $-$	\\
	& (0.907)	& V/R	& 1.77	& 2.90	& $-$	& $-$	& n/a	& $-$	\\ \hline
	&	&	&	&	&	&	&	& \\
2011.02.01	& 2455594.110	& EW [\AA]	& $-$7.63	& $-$0.51	& $-$	& $-$	& $-$0.38	& $-$	\\
	& (0.917)	& V/R	& 1.87	& 1.99	& $-$	& $-$	& 1.51	& $-$	\\ \hline
2011.02.02	& 2455595.081	& EW [\AA]	& $-$7.25	& $-$0.91	& $-$	& $-$	& $-$0.60	& $-$	\\
	& (0.926)	& V/R	& 1.90	& 2.53	& $-$	& $-$	& 1.65	& $-$	\\ \hline
2011.02.03	& 2455596.047	& EW [\AA]	& $-$7.16	& $-$0.82	& $-$	& $-$	& $-$0.32	& $-$	\\
	& (0.953)	& V/R	& 1.72	& 2.41	& $-$	& $-$	& 2.08	& $-$	\\ \hline
2011.02.04	& 2455597.097	& EW [\AA]	& $-$7.34	& $-$0.79	& $-$	& $-$	& $-$0.48	& $-$	\\
	& (0.944)	& V/R	& 1.87	& 3.07	& $-$	& $-$	& 1.77	& $-$	\\ \hline
2011.02.05	& 2455598.093	& EW [\AA]	& $-$7.56	& $-$1.00	& $-$	& $-$	& $-$0.54	& $-$	\\
	& (0.953)	& V/R	& 1.77	& 2.14	& $-$	& $-$	& 2.07	& $-$	\\ \hline
2011.02.24	& 2455616.934	& EW [\AA]	& $-$8.29	& $-$1.07	& $-$	& $-$	& $-$0.56	& $-$	\\
	& (0.124)	& V/R	& 2.06	& 1.97	& $-$	& $-$	& 1.60	& $-$	\\ \hline
2011.03.02	& 2455622.979	& EW [\AA]	& $-$8.72	& $-$0.88	& $-$	& $-$	& $-$0.73	& $-$	\\
	& (0.179)	& V/R	& 1.89	& 1.79	& $-$	& $-$	& 1.55	& $-$	\\ \hline
2011.03.10	& 2455631.042	& EW [\AA]	& $-$8.71	& $-$0.86	& $-$	& $-$	& $-$1.18	& $-$	\\
	& (0.252)	& V/R	& 2.21	& 1.96	& $-$	& $-$	& 1.67	& $-$	\\ \hline
2011.03.13	& 2455634.062	& EW [\AA]	& $-$8.30	& $-$1.42	& $-$	& $-$	& $-$1.27	& $-$	\\
	& (0.280)	& V/R	& 2.08	& 1.70	& $-$	& $-$	& 1.96	& $-$	\\ \hline
2011.03.28	& 2455649.017	& EW [\AA]	& $-$7.03	& $-$0.88	& $-$	& $-$0.48	& $-$0.74	& 0.19	\\
	& (0.415)	& V/R	& 2.37	& 2.08	& $-$	& 2.03	& 1.95	& n/a	\\ \hline
2011.09.07	& 2455812.272	& EW [\AA]	& $-$5.96	& $-$0.77	& $-$	& $-$0.54	& $-$0.50	& 0.30	\\
	& (0.896)	& V/R	& 0.55	& 0.34	& $-$	& 0.73	& 0.58	& n/a	\\ \hline
2011.09.18	& 2455823.254	& EW [\AA]	& $-$6.19	& $-$0.68	& $-$	& $-$0.68	& $-$0.68	& 0.17	\\
	& (0.996)	& V/R	& 0.59	& 0.36	& $-$	& 0.91	& 0.75	& n/a	\\ \hline
2011.09.22	& 2455827.230	& EW [\AA]	& $-$6.01	& $-$0.57	& $-$	& $-$0.39	& $-$0.56	& 0.26	\\
	& (0.032)	& V/R	& 0.56	& 0.35	& $-$	& 0.77	& 0.62	& n/a	\\ \hline
2011.09.23	& 2455828.230	& EW [\AA]	& $-$6.11	& $-$0.69	& $-$	& $-$0.40	& $-$0.56	& 0.35	\\
	& (0.041)	& V/R	& 0.58	& 0.33	& $-$	& 0.69	& 0.51	& n/a	\\ \hline
2011.09.24	& 2455829.231	& EW [\AA]	& $-$6.07	& $-$0.60	& $-$	& $-$0.35	& $-$0.57	& 0.33	\\
	& (0.050)	& V/R	& 0.58	& 0.35	& $-$	& 0.82	& 0.63	& n/a	\\ \hline
2011.09.26	& 2455831.269	& EW [\AA]	& $-$5.82	& $-$0.67	& $-$	& $-$0.53	& $-$0.63	& 0.14	\\
	& (0.069)	& V/R	& 0.57	& 0.48	& $-$	& 0.98	& 0.74	& n/a	\\ \hline
2011.10.07	& 2455842.173	& EW [\AA]	& $-$6.15	& $-$0.67	& $-$	& $-$0.53	& $-$0.63	& 0.41	\\
	& (0.168)	& V/R	& 0.59	& 0.49	& $-$	& n/a	& 0.94	& n/a	\\ \hline
	&	&	&	&	&	&	&	& \\
2011.12.10	& 2455906.193	& EW [\AA]	& $-$5.01	& $-$0.38	& $-$	& $-$0.02	& $-$0.14	& 0.34	\\
	& (0.748)	& V/R	& 0.76	& 0.86	& $-$	& 2.28	& 1.99	& n/a	\\ \hline
2011.12.30	& 2455926.269	& EW [\AA]	& $-$5.92	& $-$0.68	& $-$	& $-$0.08	& $-$0.27	& 0.23	\\
	& (0.930)	& V/R	& 0.96	& 1.24	& $-$	& 2.19	& 1.89	& n/a	\\ \hline
2012.01.06	& 2455933.096	& EW [\AA]	& $-$6.19	& $-$0.11	& $-$	& $-$	& $\times$	& $-$	\\
	& (0.992)	& V/R	& 0.91	& 1.03	& $-$	& $-$	& $\times$	& $-$	\\ \hline
2012.01.07	& 2455934.225	& EW [\AA]	& $-$6.24	& $-$1.28	& $-$	& $-$	& $\times$	& $-$	\\
	& (0.003)	& V/R	& 0.99	& 1.04	& $-$	& $-$	& $\times$	& $-$	\\ \hline
2012.01.08	& 2455935.155	& EW [\AA]	& $-$5.84	& $-$0.64	& $-$	& $-$	& $\times$	& $-$	\\
	& (0.011)	& V/R	& 0.95	& 1.18	& $-$	& $-$	& $\times$	& $-$	\\ \hline
2012.02.11	& 2455969.152	& EW [\AA]	& $-$5.89	& $-$0.67	& $-$	& $-$0.15	& $-$0.33	& 0.25	\\
	& (0.319)	& V/R	& 1.23	& 1.39	& $-$	& 1.80	& 1.56	& n/a	\\ \hline
2012.03.27	& 2456013.958	& EW [\AA]	& $-$4.95	& $-$0.51	& $-$	& $-$0.01	& $-$0.16	& 0.23	\\
	& (0.726)	& V/R	& 1.34	& 1.61	& $-$	& 1.53	& 2.15	& n/a	\\ \hline
2012.04.26	& 2456043.965	& EW [\AA]	& $-$5.63	& $-$0.81	& $-$	& $-$0.03	& $-$0.22	& 0.06	\\
	& (0.998)	& V/R	& 1.17	& 0.92	& $-$	& 0.83	& 0.90	& n/a	\\ \hline
\end{longtable}

\begin{longtable}[]{*{4}{c}||*{4}{c}}
	\caption{
Wing position of the H$\alpha$ line profile ($\mathrm{km\;s^{-1}}$).
Columns 1, 5 and 2, 6 list the observational dates and HJD, respectively.
The violet wing position is shown in columns 3 and 7, while the red wing position in alums 4 and 8.
}
	\label{tbl:Obs_wing}\\
	\hline
	 &	& \multicolumn{2}{c||}{$RV_{\mathrm{wing}}$ [$\mathrm{km\;s^{-1}}$]}	&	&	& \multicolumn{2}{c}{$RV_{\mathrm{wing}}$ [$\mathrm{km\;s^{-1}}$]} \\
	 Date	& HJD	& B	& R	& Date	& HJD	& B	& R \\
	\hline \hline
	\endhead
	\hline
	\endfoot
	\hline
	\endlastfoot
	2005.11.24	& 2453699.060	& $-$435	& 523	& 2008.12.29	& 2454830.165	& $-$383	& 541 \\
	2005.11.25	& 2453700.060	& $-$449	& 529	& 2008.12.30	& 2454831.157	& $-$423	& 561 \\
	2005.11.27	& 2453702.060	& $-$463	& 542	& 2008.12.31	& 2454832.112	& $-$355	& 517 \\
	2005.11.29	& 2453704.031	& $-$473	& 554	& 2009.01.01	& 2454833.008	& $-$333	& 517 \\
	2005.11.30	& 2453705.081	& $-$479	& 524	& 2009.01.02	& 2454834.021	& $-$325	& 505 \\
	2005.12.01	& 2453706.075	& $-$468	& 529	& 2009.01.03	& 2454835.028	& $-$359	& 516 \\
	2005.12.03	& 2453708.651	& $-$543	& 560	& 2009.01.05	& 2454837.114	& $-$448	& 555 \\
	2006.12.18	& 2454088.100	& $-$619	& 591	& 2009.01.06	& 2454838.138	& $-$319	& 510 \\
	2007.11.07	& 2454412.184	& $-$511	& 534	& 2009.01.07	& 2454839.180	& $-$414	& 546 \\
	2007.11.08	& 2454413.172	& $-$479	& 492	& 2009.01.08	& 2454840.194	& $-$364	& 542 \\
	2007.11.09	& 2454414.141	& $-$456	& 527	& 2009.01.10	& 2454842.186	& $-$331	& 532 \\
	2007.11.10	& 2454415.232	& $-$407	& 479	& 2009.01.12	& 2454844.157	& $-$399	& 560 \\
	2007.11.11	& 2454416.166	& $-$497	& 528	& 2009.03.12	& 2454902.965	& $-$293	& 560 \\
	2007.11.13	& 2454418.159	& $-$462	& 534	& 2009.04.09	& 2454930.955	& $-$313	& 542 \\
	2007.11.14	& 2454419.238	& $-$413	& 515	& 2009.08.31	& 2455075.264	& $-$321	& 447 \\
	2007.12.16	& 2454451.168	& $-$415	& 508	& 2009.09.13	& 2455088.242	& $-$387	& 526 \\
	2007.12.19	& 2454454.294	& $-$452	& 536	& 2009.10.18	& 2455123.143	& $-$405	& 535 \\
	2007.12.20	& 2454455.264	& $-$434	& 518	& 2009.11.05	& 2455141.131	& $-$393	& 458 \\
	2007.12.26	& 2454461.090	& $-$455	& 499	& 2009.11.14	& 2455150.289	& $-$385	& 517 \\
	2008.01.02	& 2454468.022	& $-$500	& 575	& 2009.11.19	& 2455155.255	& $-$403	& 567 \\
	2008.01.31	& 2454497.004	& $-$440	& 573	& 2009.11.23	& 2455159.301	& $-$373	& 520 \\
	2008.03.11	& 2454537.002	& $-$552	& 571	& 2009.11.30	& 2455166.233	& $-$374	& 499 \\
	2008.03.21	& 2454547.023	& $-$425	& 578	& 2009.12.01	& 2455167.084	& $-$427	& 530 \\
	2008.10.01	& 2454741.208	& $-$563	& 561	& 2009.12.03	& 2455169.277	& $-$402	& 501 \\
	2008.10.11	& 2454751.179	& $-$472	& 539	& 2009.12.04	& 2455170.229	& $-$385	& 498 \\
	2008.10.15	& 2454755.186	& $-$464	& 570	& 2009.12.05	& 2455171.088	& $-$430	& 525 \\
	2008.10.19	& 2454759.141	& $-$509	& 535	& 2009.12.06	& 2455172.033	& $-$413	& 531 \\
	2008.11.04	& 2454775.102	& $-$417	& 580	& 2009.12.07	& 2455173.311	& $-$381	& 515 \\
	2008.12.12	& 2454813.068	& $-$346	& 531	& 2009.12.08	& 2455174.150	& $-$407	& 533 \\
	2008.12.25	& 2454826.207	& $-$406	& 505	& 2009.12.09	& 2455175.060	& $-$336	& 525 \\
	2008.12.26	& 2454827.097	& $-$399	& 532	& 2009.12.12	& 2455178.078	& $-$382	& 520 \\
	2008.12.27	& 2454828.129	& $-$392	& 540	& 2009.12.16	& 2455181.987	& $-$400	& 533 \\
	2008.12.28	& 2454829.187	& $-$403	& 549	& 2009.12.19	& 2455185.060	& $-$408	& 562 \\
	2009.12.22	& 2455188.223	& $-$415	& 550	& 2011.01.30	& 2455592.110	& $-$268	& 535 \\
	2009.12.23	& 2455189.126	& $-$345	& 551	& 2011.01.31	& 2455592.992	& $-$196	& 483 \\
	2009.12.26	& 2455192.903	& $-$361	& 521	& 2011.02.01	& 2455594.110	& $-$256	& 516 \\
	2010.01.03	& 2455200.104	& $-$343	& 505	& 2011.02.02	& 2455595.081	& $-$244	& 484 \\
	2010.01.06	& 2455202.960	& $-$382	& 537	& 2011.02.03	& 2455596.047	& $-$188	& 507 \\
	2010.01.13	& 2455210.120	& $-$335	& 470	& 2011.02.04	& 2455597.097	& $-$251	& 497 \\
	2010.01.16	& 2455213.053	& $-$356	& 546	& 2011.02.05	& 2455598.093	& $-$256	& 512 \\
	2010.01.22	& 2455219.040	& $-$406	& 522	& 2011.02.24	& 2455616.934	& $-$245	& 533 \\
	2010.01.26	& 2455223.151	& $-$333	& 502	& 2011.03.02	& 2455622.979	& $-$239	& 495 \\
	2010.02.03	& 2455231.072	& $-$331	& 536	& 2011.03.10	& 2455631.042	& $-$249	& 522 \\
	2010.02.04	& 2455232.133	& $-$314	& 481	& 2011.03.13	& 2455634.062	& $-$233	& 524 \\
	2010.02.19	& 2455247.045	& $-$277	& 532	& 2011.03.28	& 2455649.017	& $-$196	& 483 \\
	2010.02.24	& 2455251.965	& $-$335	& 535	& 2011.09.07	& 2455812.272	& $-$308	& 364 \\
	2010.02.28	& 2455256.035	& $-$280	& 431	& 2011.09.18	& 2455823.254	& $-$302	& 390 \\
	2010.03.17	& 2455272.960	& $-$255	& 510	& 2011.09.22	& 2455827.230	& $-$302	& 378 \\
	2010.03.26	& 2455281.982	& $-$294	& 514	& 2011.09.23	& 2455828.230	& $-$304	& 412 \\
	2010.03.30	& 2455285.925	& $-$205	& 407	& 2011.09.24	& 2455829.231	& $-$296	& 489 \\
	2010.08.20	& 2455429.309	& $-$412	& 371	& 2011.09.26	& 2455831.269	& $-$306	& 366 \\
	2010.09.16	& 2455456.298	& $-$272	& 424	& 2011.10.07	& 2455842.173	& $-$318	& 392 \\
	2010.10.12	& 2455482.212	& $-$265	& 358	& 2011.12.10	& 2455906.193	& $-$338	& 543 \\
	2010.10.22	& 2455492.219	& $-$202	& 402	& 2011.12.30	& 2455926.269	& $-$337	& 533 \\
	2010.10.25	& 2455495.328	& $-$249	& 484	& 2012.01.06	& 2455933.096	& $-$335	& 563 \\
	2010.11.03	& 2455504.075	& $-$332	& 442	& 2012.01.07	& 2455934.225	& $-$297	& 576 \\
	2010.11.04	& 2455505.187	& $-$332	& 452	& 2012.01.08	& 2455935.155	& $-$268	& 562 \\
	2010.11.18	& 2455519.187	& $-$369	& 497	& 2012.02.11	& 2455969.152	& $-$281	& 556 \\
	2010.12.14	& 2455544.966	& $-$257	& 479	& 2012.03.27	& 2456013.958	& $-$242	& 535 \\
	2011.01.27	& 2455589.165	& $-$278	& 468	& 2012.04.26	& 2456043.965	& $-$261	& 541 \\
	2011.01.29	& 2455591.075	& $-$245	& 519 \\
\end{longtable}

\end{document}